\documentclass[longauth,letter]{aa} % for the letters 
\usepackage[dvipsnames,svgnames]{xcolor}
\usepackage{graphicx}
\usepackage{subcaption}
\usepackage{lscape}
\usepackage{longtable}
\usepackage{multirow}
\usepackage{gensymb}
\usepackage{adjustbox}
\usepackage{amsmath}
\usepackage{ragged2e}
\usepackage{amssymb}
\usepackage{rotating}
\usepackage{xspace}
\usepackage{makecell}
\usepackage[comma,authoryear]{natbib}
\usepackage[colorlinks=true,allcolors=Blue]{hyperref}

\usepackage{txfonts}
\newcommand*{\Mjup}{\ensuremath{M_{\textrm{Jup}}}\xspace}
\newcommand*{\mjup}{\ensuremath{M_{\textrm {Jup}}}\xspace}

\newcommand*{\Msun}{\ensuremath{M_\odot}}
\newcommand*{\msun}{\ensuremath{M_\odot}}
\newcommand*{\Lsun}{\ensuremath{L_\odot}}

\newcommand*{\ebv}{\ensuremath{\mathrm{E(B-V)}}}
\newcommand*{\teff}{\ensuremath{T_{\mathrm{eff}}}}

\begin{document}
   \title{An imaged $15 \mjup$ companion within a hierarchical quadruple system\thanks{Based on data obtained with the ESO/VLT SPHERE instrument under programs 1101.C-0258(A/E).}
}
    \author{A. Chomez\inst{\ref{LESIA},\ref{IPAG}} 
    \and V. Squicciarini\inst{\ref{LESIA},\ref{INAF}} 
    \and A.-M. Lagrange\inst{\ref{LESIA},\ref{IPAG}} 
    \and P. Delorme\inst{\ref{IPAG}} 
    \and G. Viswanath \inst{\ref{IfA}}
    \and M. Janson \inst{\ref{IfA}}
    \and O. Flasseur\inst{\ref{CRAL}} 
    \and G. Chauvin\inst{\ref{OCA}} 
    \and M. Langlois\inst{\ref{CRAL}} 
    \and P. Rubini\inst{\ref{PIXYL}} 
    \and S. Bergeon\inst{\ref{IPAG}}
    \and D. Albert\inst{\ref{OSUG}}
    \and M. Bonnefoy \inst{\ref{IPAG}}    
    \and S. Desidera \inst{\ref{INAF}}
    \and N. Engler \inst{\ref{ETH_Zurich}} 
    \and R. Gratton \inst{\ref{INAF}} 
    \and T. Henning \inst{\ref{mpia}}
    \and E.E. Mamajek \inst{\ref{JPL}}
    \and G.-D. Marleau \inst{\ref{duisburg},\ref{tuebingen},\ref{bern},\ref{mpia}}
    \and M.R. Meyer \inst{\ref{UM}} 
    \and S. Reffert \inst{\ref{Heid}}
    \and S.C. Ringqvist \inst{\ref{IfA}}
    \and M. Samland \inst{\ref{IfA}} 
    }

\institute{
\label{LESIA}
LESIA, Observatoire de Paris, Universit\'{e} PSL, CNRS, 5 Place Jules Janssen, 92190 Meudon, France \\
\email{antoine.chomez@obspm.fr}
\and
\label{IPAG}
Univ. Grenoble Alpes, CNRS-INSU, Institut de Planetologie et d'Astrophysique de Grenoble (IPAG) UMR 5274, Grenoble, F-38041, France;
\and
\label{INAF}
INAF – Osservatorio Astronomico di Padova; Vicolo
dell’Osservatorio 5, I-35122 Padova, Italy
\and
\label{IfA}
Institutionen för astronomi, Stockholms universitet; AlbaNova universitetscentrum, SE-106 91 Stockholm, Sweden
\and
\label{CRAL}
Univ. Lyon, Univ. Lyon1, ENS de Lyon, CNRS, Centre de Recherche Astrophysique de Lyon (CRAL) UMR5574, F-69230 Saint-Genis-Laval, France
\and
\label{OCA}
Université Côte d’Azur, OCA, CNRS, Lagrange, France
\and
\label{PIXYL}
Pixyl S.A. La Tronche, France
\and
\label{ETH_Zurich}
ETH Zurich, Institute for Particle Physics and Astrophysics, Wolfgang-Pauli-Strasse 27, CH-8093 Zurich, Switzerland 
\and
\label{mpia}
Max-Planck-Institut f\"ur Astronomie, K\"onigstuhl 17, 69117 Heidelberg, Germany
\and
\label{JPL}
Jet Propulsion Laboratory, California Institute of Technology; 4800 Oak Grove Drive, Pasadena CA 91109, USA
\and
\label{duisburg}
Fakult\"at f\"ur Physik, Universit\"at Duisburg-Essen, Lotharstra\ss{}e 1, 47057 Duisburg, Germany
\and
\label{tuebingen}
Instit\"ut f\"ur Astronomie und Astrophysik, Universit\"at T\"ubingen, Auf der Morgenstelle 10, 72076 T\"ubingen, Germany
\and
\label{bern}
Physikalisches Institut, Universit\"at Bern, Gesellschaftsstr.~6, 3012 Bern, Switzerland
\and
\label{UM}
Department of Astronomy, University of Michigan; 1085 S. University Ave, Ann Arbor MI 48109, USA
\and
\label{Heid}
Landessternwarte, Zentrum für Astronomie der Universität Heidelberg; Königstuhl 12, 69117 Heidelberg, Germany
\and
\label{OSUG}
Université Grenoble Alpes, CNRS, Observatoire des Sciences de l’Univers de Grenoble (OSUG), Grenoble, France
}

   \date{\today}

   % \abstract{}{}{}{}{}
   % 5 {} token are mandatory
   
   \abstract
   % context heading (optional)
   {
    Since 2019, the direct imaging B-star Exoplanet Abundance Study (BEAST) at SPHERE@VLT has been scanning the surroundings of young B-type stars in order to ascertain the ultimate frontiers of giant planet formation. Recently, the $17^{+3}_{-4}$ Myr HIP~81208 was found to host a close-in ($\sim 50$ au) brown dwarf and a wider ($\sim 230$ au) late M star around the central $2.6 \msun$ primary.
   } %leave it empty if necessary
   {
   Alongside the continuation of the survey, we are undertaking a complete reanalysis of archival data aimed at improving detection performances so as to uncover additional low-mass companions.
   }
   % aims heading (mandatory)
   {
   We present here a new reduction of the observations of HIP~81208 using PACO ASDI, a recent and powerful algorithm dedicated to processing high-contrast imaging datasets, as well as more classical algorithms and a dedicated PSF-subtraction approach. The combination of different techniques allowed for a reliable extraction of astrometric and photometric parameters.
   }
   % methods heading (mandatory)
   {
   A previously undetected source was recovered at a short separation from the C component of the system. Proper motion analysis provided robust evidence for the gravitational bond of the object to HIP~81208~C. Orbiting C at a distance of $\sim 20$ au, this $15~\Mjup$ brown dwarf becomes the fourth object of the hierarchical HIP~81208 system.
   }
   % results heading (mandatory)
   {
   Among the several BEAST stars which are being found to host substellar companions, HIP~81208 stands out as a particularly striking system. As the first stellar binary system with substellar companions around each component ever found by direct imaging, it yields exquisite opportunities for thorough formation and dynamical follow-up studies.
    }  


\titlerunning{An imaged $15 \mjup$ companion within a hierarchical quadruple system}

\maketitle

\section{Introduction}

The formation of planets in binary systems, and chiefly the tightest ($\lesssim 50$ au) ones, is a vibrant subject in exoplanetology. Indeed, binary systems are complex environments from a dynamical point of view, severely affecting the size of protoplanetary disks and their capability to either form massive enough cores to undergo runaway gas accretion \citep[core accretion (CA);][]{CA_intro} or induce low enough Toomre Q values to trigger gravitational instability \citep[GI;][]{GI_intro}. Whether substellar companions can form critically depends on the stars properties, their physical separations, and the disk initial properties \citep{binary_disks,Jang_Condell_2015,Silsbee_2021}. 

From an observational standpoint, $\sim 300$ S-type substellar companions (companions orbiting one component of a binary system) within binary systems are known to date \citep{fontanive21,chauvin22} -- their frequency being anti-correlated with binary separation \citep{wang14} --, as well as triple and higher-order planet-hosting systems in strongly hierarchical configurations \citep{roberts15,cuntz22}. Indirect techniques have identified a few systems where both components host substellar companions \citep[see, e.g.,][]{desidera14,lissauer14,udry19}; notably, HD 41004 stands out due to its close A-B separation ($\sim 23$ au): with a $m \sin{i} = 18~\Mjup$ brown dwarf orbiting at 0.017 au from component B ($0.4~\Msun$) and a $m \sin{i} = 2.5~\Mjup$ companion around component A ($0.7~\Msun$) on an orbit with semi-major axis $a=1.6$ au and a large eccentricity $e=0.4$ \citep{Zucker_2003,Zucker_2004}.

In the course of a new analysis of archival data obtained through the SPHERE high-contrast imager \citep{Beuzit_sphere}, we detected a new companion in the young ($17^{+3}_{-4}$ Myr) HIP~81208 system. HIP~81208 was observed as part of the BEAST survey dedicated to the search for exoplanets around 85 B-type members of the Scorpius Centaurus (Sco-Cen) association \citep{Janson_beast}. Located in the Upper Centaurus-Lupus (UCL) subgroup of Sco-Cen at a distance of $148.7^{+1.5}_{-1.3}$~pc \citep{gaia_dr3}, it has been recently identified as a triple system, where the A component is a $2.58 \pm 0.06$~\Msun~B9V star, the B component is a $67^{+6}_{-7}~\Mjup$ brown dwarf orbiting HIP~81208~A at about 50 au, and the C component is a low-mass star of $0.135^{+0.010}_{-0.013}$~\Msun ~orbiting HIP~81208~A at about 230 au \citep{Viswanath_hip81208_tmp}. 
The newly found companion \citep[hereafter Cb, following naming conventions for hierarchical systems;][]{star_nomenclature,planet_nomenclature} is orbiting the C component, making HIP~81208 the first binary system with substellar companions around both components ever discovered through DI.

We present SPHERE data and data reduction in Sect.~\ref{sec:data_analysis}; after confirming the bound nature of the companion candidate, we describe its properties in Sect.~\ref{sec:Cb_properties}. A discussion on the peculiar properties of this quadruple system follows in Sect.~\ref{sec:conclusion}.

%==========================================================================
\section{Data analysis }
\label{sec:data_analysis}
%=============================================
%
\subsection{SPHERE data}

HIP~81208 was observed twice by SPHERE \citep{Beuzit_sphere} on August 6, 2019 and on April 5, 2022.
Both observations were conducted using the telescope in pupil-stabilized mode. This allow the use of angular and spectral differential imaging \citep[ASDI,][]{Marois_adi} post processing techniques.
In each case, an unsaturated, non coronagraphic image (point spread function; PSF) of the primary was obtained for flux calibration purposes, as well as a coronagraphic exposure with a waffle pattern applied to the mirror \citep{Cantalloube_2019}, for centering purposes, before and after the main coronagraphic exposures. The N-ALC-YJH-S coronagraph was used, allowing the infrared dual-band imager and spectrograph \citep[IRDIS,][]{Dohlen_irdis} to observe in the K1 and K2 bands while the integral field spectrograph \citep[IFS,][]{Claudi_ifs} observed in the YJH bands. Because the source of interest for this letter is outside the field of view of IFS, only IRDIS data will be considered hereafter. Table \ref{tab:obs_cond_table} summarizes the observing conditions as well as the setup for the two observations, the same already used in \citep{Viswanath_hip81208_tmp}.

\begin{table}[t!]
    \centering
    \begin{tabular}{l|c|c}
            & \makecell{First epoch \\ (2019-08-06)} & \makecell{Second epoch \\ (2022-04-05)}  \\
            \hline
            \hline
        IRDIS filter & DB\_K12 & DB\_K12 \\
        DIT(s)$\times$Nframe & 64$\times$48 & 64$\times$48\\
        $\Delta$PA ($\degree$) & 59.09 & 57.33 \\
        seeing (")$^a$ & 0.64 & 0.57 \\
        Airmass$^a$ & 1.020 & 1.020 \\
        $\tau_0$ (ms)$^{a}$ & 8.1 & 5.3 \\
        Program ID & 1101.C-0258(A) & 1101.C-0258(E) \\
        \hline
        \hline
    \end{tabular}
    \caption{Observation logs for the two epochs.}
    \label{tab:obs_cond_table}
    \vspace{0.2cm}
    \RaggedRight
    \textbf{Notes:} DIT = detector integration time per frame, $\Delta$PA = amplitude of the parallactic rotation, $\tau_0$ = coherence time. $^a$: values extracted from the updated DIMM info and averaged over the sequence.
\end{table}

\subsection{Data reduction}
\label{sec:data_reduction}

Data reduction was performed on the COBREX Data Center, a modified and improved server based on the High Contrast Data Center (HC-DC, formerly SPHERE Data Center), \cite{Delorme_sphereDC} and aimed at improving detection capabilities with existing SPHERE images by means of the \verb+PACO+ algorithm. More specifically, we used \verb+PACO ASDI+ \citep{Flasseur_paco, flasseur2020robustness, Flasseur_asdi} as well as the No-ADI routine embedded in the \verb+SPECAL+ software \citep{Galicher_specal} as post-processing algorithms.
The pre-reduction pipeline (i.e. going from raw data to calibrated 4D datacube) is identical to the one implemented in the HC-DC, performing dark, flat, distortion and bad pixel corrections.

PACO is modelling the noise using a multi Gaussian model at a local scale on small patches, allowing a better estimation of the temporal and spectral correlation of the noise.
The full details on the improvements of the pre-reduction pipeline as well as the optimization regarding the ASDI mode of \verb+PACO+, and the obtained performances are described in \citet{chomez2023preparing}. \verb+PACO+ provides a contrast gain between 1 and 2 magnitudes at all separations as well as reliable and statistically grounded signal to noise (S/N) detection and contrast maps in an unsupervised fashion compared to more classical algorithms like TLOCI-ADI used in \verb+SPECAL+ \citep{Galicher_specal}. 

\begin{figure*}[t!]
    \centering
    \begin{subfigure}[b]{0.475\textwidth}
        \centering
        \includegraphics[width=\textwidth]{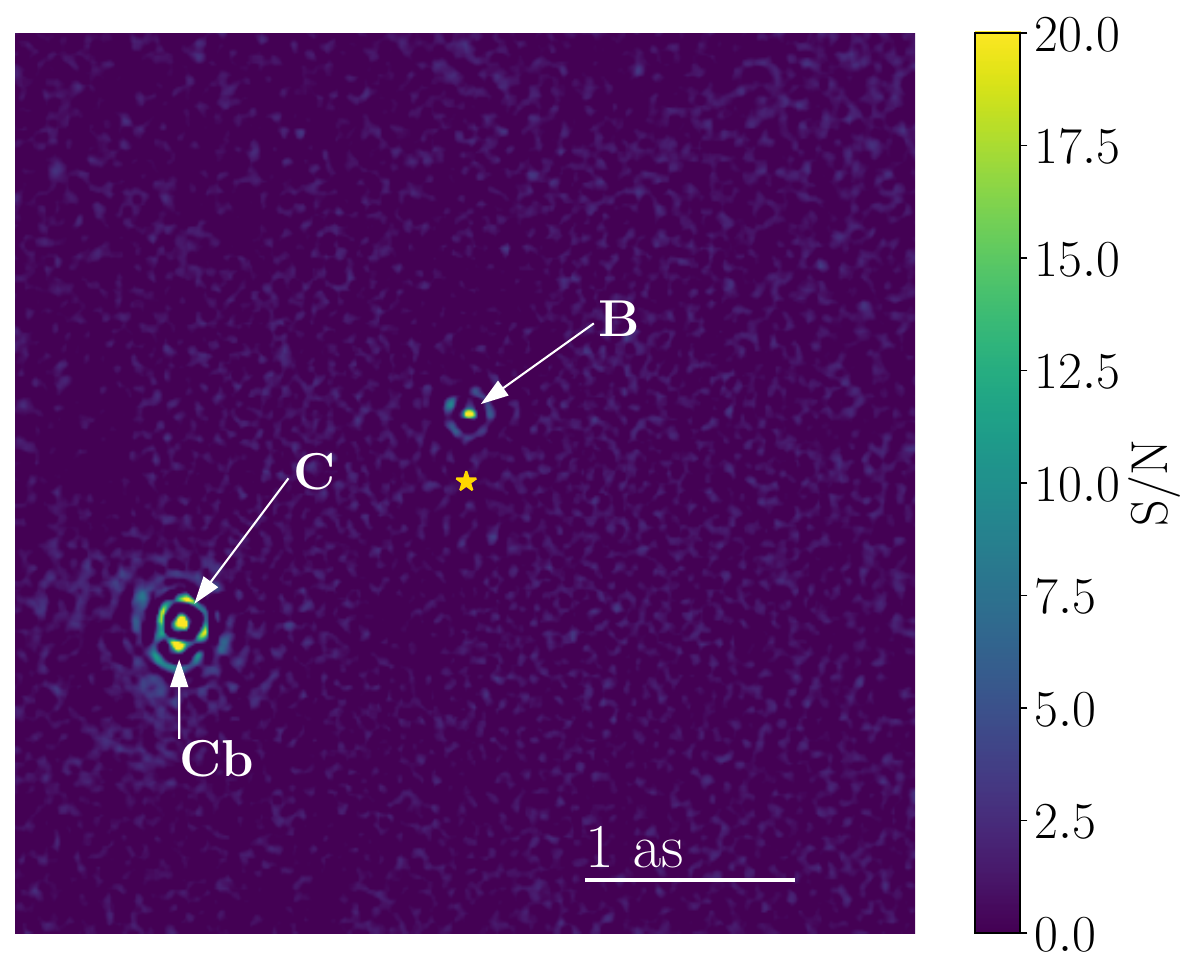}
        %\caption[]%
        %{{\small 2019 PACO S/N map}}    
        \label{subfig:paco_2019}
    \end{subfigure}
    \hfill
    \begin{subfigure}[b]{0.475\textwidth}  
        \centering 
        \includegraphics[width=\textwidth]{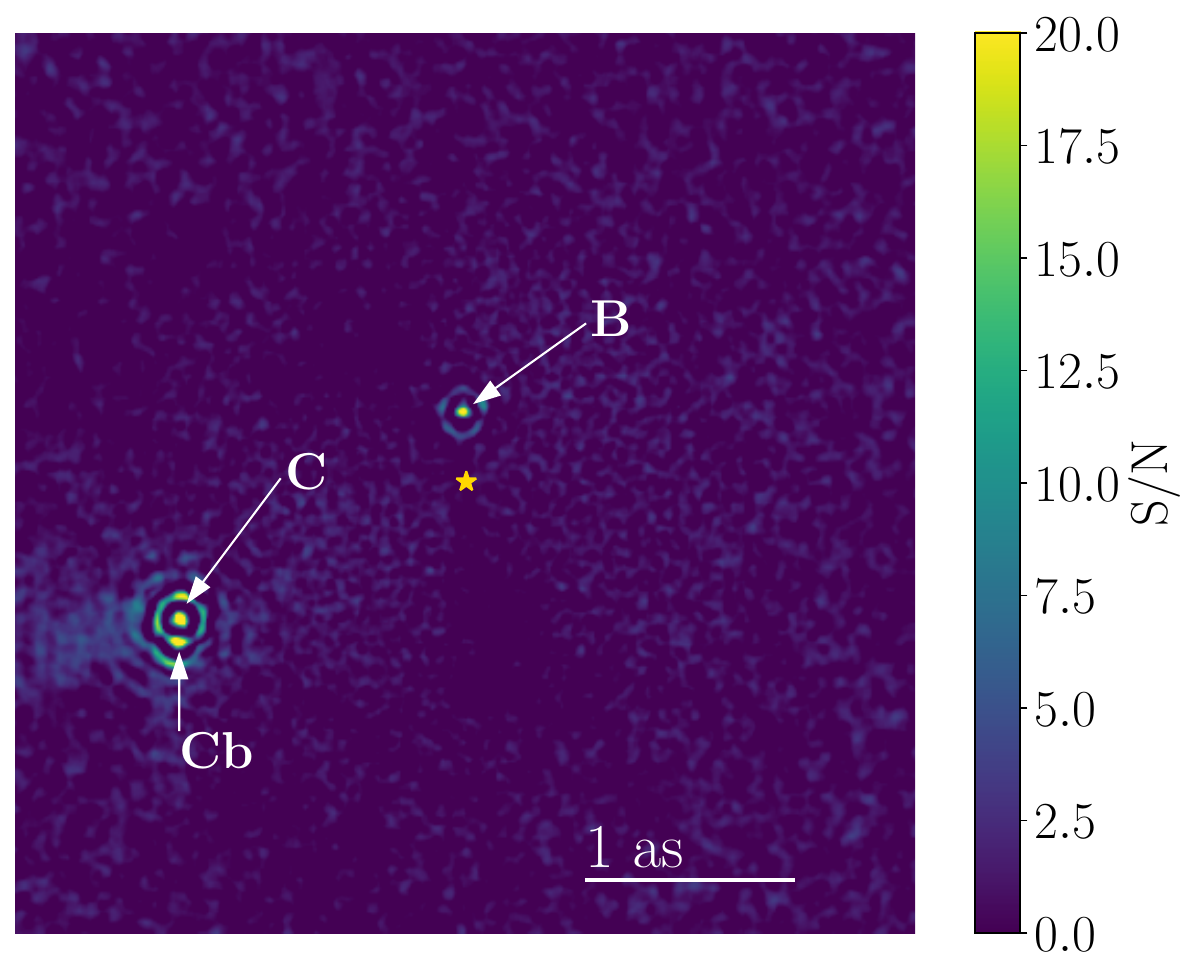}
        %\caption[]%
        %{{\small 2022 PACO S/N map}}    
        \label{subfig:paco_2022}
    \end{subfigure}
    \vskip\baselineskip
    \begin{subfigure}[b]{0.475\textwidth}   
        \centering 
        \includegraphics[width=\textwidth]{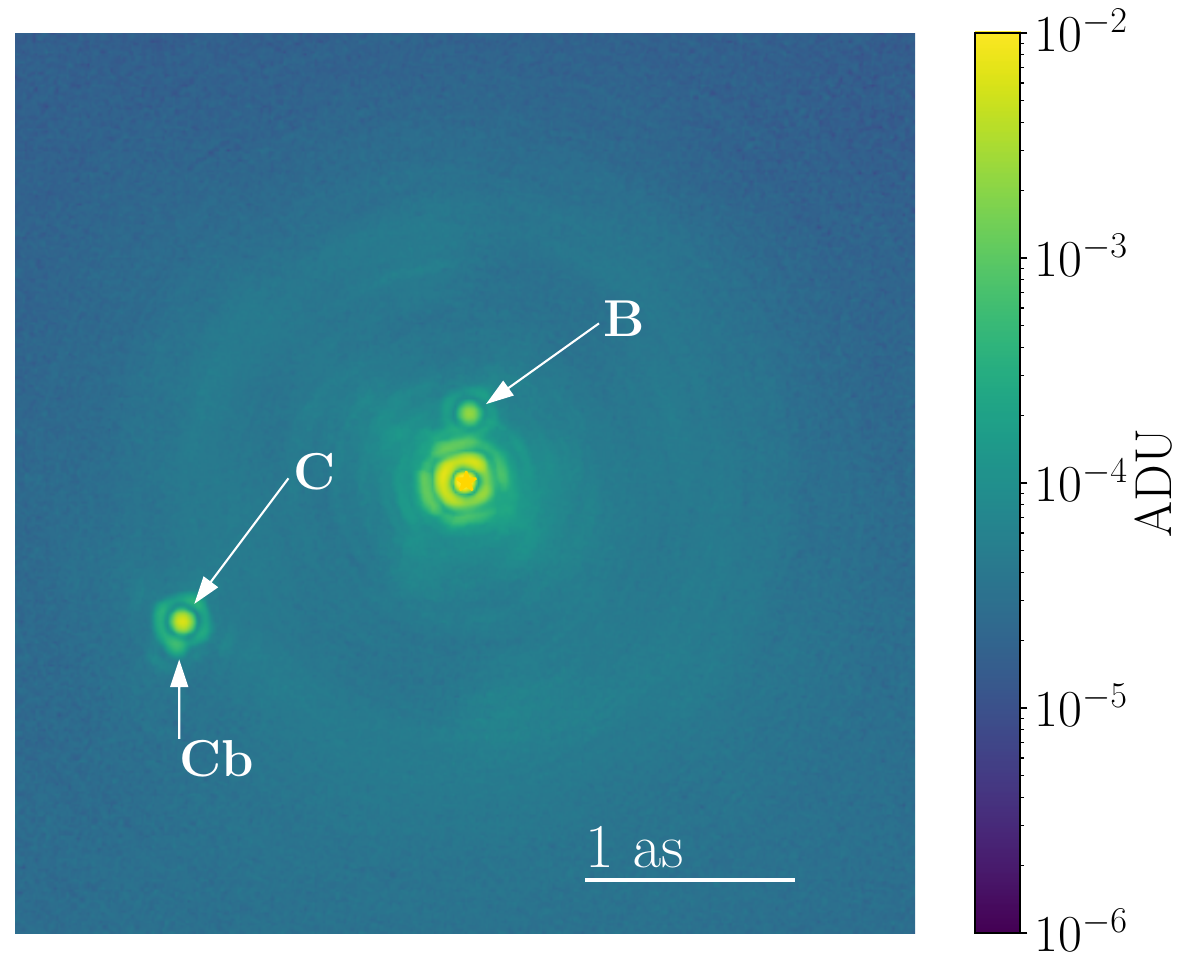}
        %\caption[]%
        %{{\small 2019 No-ADI residual map}}    
        \label{subfig:nadi_2019}
    \end{subfigure}
    \hfill
    \begin{subfigure}[b]{0.475\textwidth}   
        \centering 
        \includegraphics[width=\textwidth]{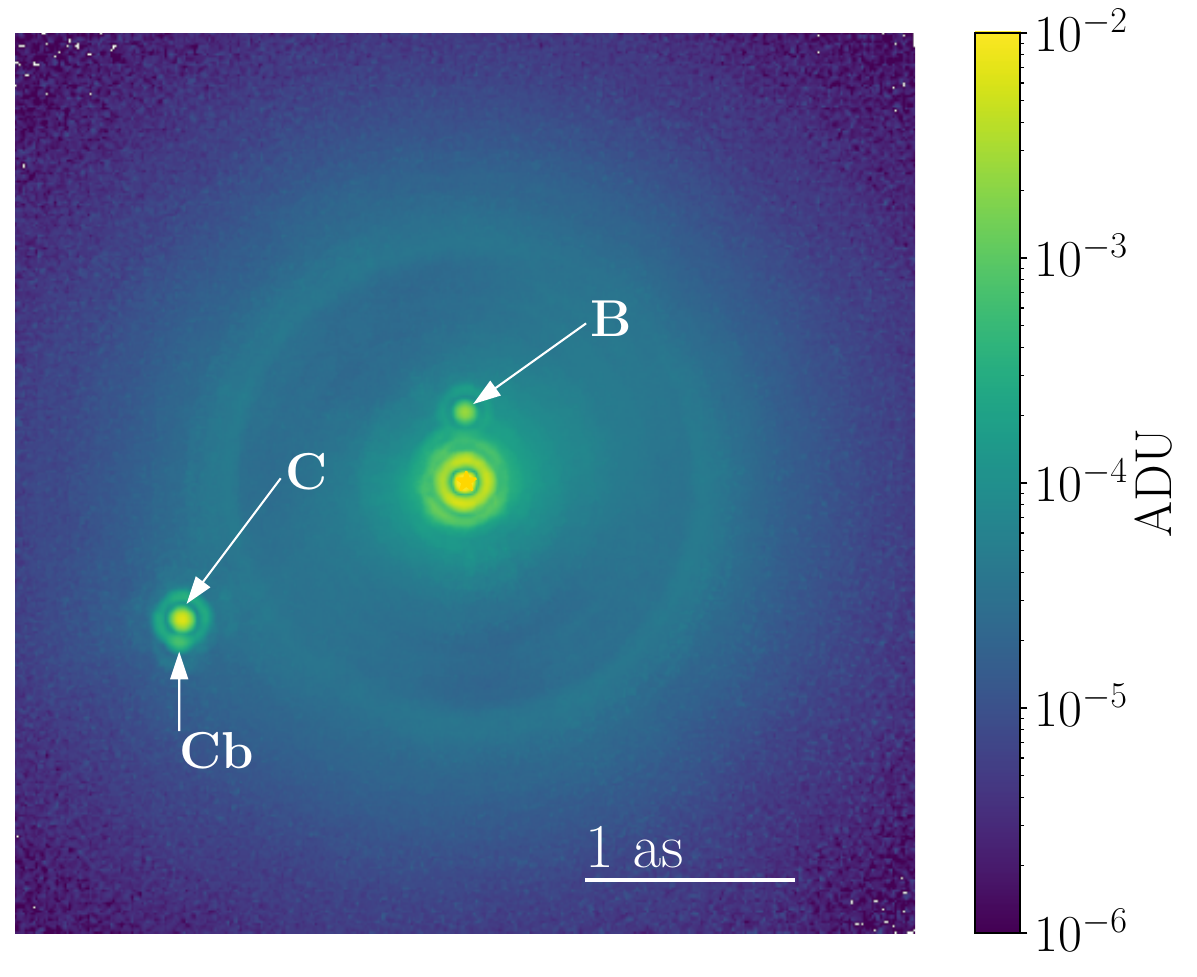}
        %\caption[]%
        %{{\small 2022 No-ADI residual map}}    
        \label{subfig:nadi_2022}
    \end{subfigure}
    \caption{S/N maps produced by PACO (top row) and residual maps produced by SPECAL no-ADI (bottom row) for both epochs (2019 on the left and 2022 on the right). The B and C components are retrieved and the reductions unveil the additional Cb companion. The central star (hidden behind the coronagraphic mask) is indicated by the yellow star. } 
    \label{fig:det_maps}
\end{figure*}

Prompted by the results emerging from our \verb+PACO+ and No-ADI reductions (Sect. \ref{subsec:det_sources}), we additionally  developed a custom PSF subtraction routine, building a local PSF model to remove C and enhance detection capabilities in its immediate surroundings ( See detailed description in Appendix~\ref{Appendix_PSF}).

\begin{figure}[t!]
    \centering
    \includegraphics[width=\linewidth]{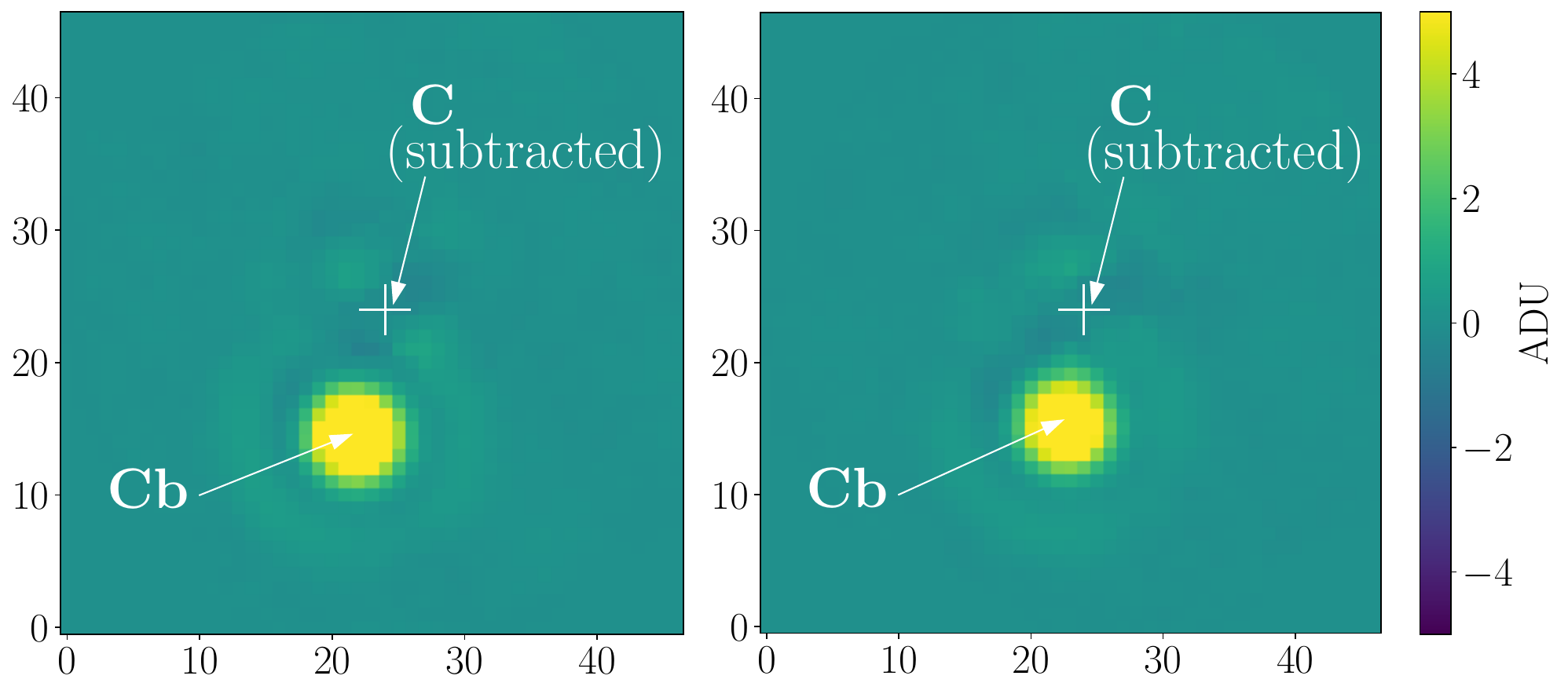}
    \caption{Zoom on the C - Cb pair after subtracting the PSF of C (2019 on the left, 2022 on the right) for the K1 band. The position of C (subtracted) is highlighted by the white cross. Cb is clearly visible after the removal of C.}
    \label{fig:psf_subtract_zoom_cb}
\end{figure}

As regards the true North, the pixel scale, and the pupil offset, we adopted the long-term IRDIS calibrations by \citet{Maire_calib}: a pixel scale of $12.258 \pm 0.004$ mas/pixel (K1 band) and $12.253 \pm 0.003$ mas/pixel (K2 band), a true North orientation of $(-1.77 \pm 0.04)^{\circ}$, and a pupil offset of $(136 \pm 0.03)^{\circ}$.

\subsection{Detected sources}
\label{subsec:det_sources}

Figure \ref{fig:det_maps} shows the S/N maps of \verb+PACO+ and residual No-ADI maps for both epochs: a new source (hereafter Cb) was detected in close proximity ($\sim 120$ mas) to C.

\verb+PACO+ detects Cb with a high S/N (26.2 in 2019, 16.9 in 2022), placing a high confidence level on the detection. Although Cb is visible by eye in the no-ADI reduction, no reliable measurement could be extracted for the source because of strong contamination by C; \verb+SPECAL+ does not provide tools to deconvolve sources. Conversely, \verb+PACO+ includes a \textit{cleaning} option designed for the case: it removes the contribution of C while characterizing Cb, enabling extraction of a reliable photometry \citep{Flasseur_asdi}. We attribute the previous non-detection of Cb to self-subtraction artefacts introduced by TLOCI -- the baseline algorithm used to process IRDIS observations in BEAST's standard reduction pipeline -- near C\footnote{Any ADI-based algorithm with a subtraction step (e.g., which does not fit the planet and the systematics simultaneously) is also suffering from this self-subtraction effect.}. 

Our PSF subtraction routine, specifically designed to investigate the surroundings of C, enabled us to solidly reveal Cb and its first Airy ring on almost 360° (see Fig. \ref{fig:psf_subtract_zoom_cb}) at both epochs. Figure \ref{fig:residual_custom_psf_sub} shows the residuals after removing both C and Cb. The highest residuals in both epochs are barely above the local background noise in each of the 48 individual frames, allowing us to robustly determine position, contrast and associated uncertainties by deriving the mean and standard deviation of the resulting 48 independent measurements. 
Notably, the source is not only visible with both algorithms at both epoch but also in each raw frame, before any post-processing algorithm is applied (Fig.~\ref{fig:frame_sequence}); unlike the nearby $1^{\text{st}}$ Airy ring of C, its separation from C does not vary with wavelength, while its rotation around C during the ADI sequence is consistent with the expected motion of a physical source (see Appendix \ref{Appendix_raw_frames}), thus ensuring it is not an artefact.

As an additional check, we also characterized B and C, finding results compatible within 1$\sigma$ to those presented by \citet{Viswanath_hip81208_tmp}\footnote{We attribute the larger astrometric uncertainties emerging in our analysis to the fact that the previous analysis did not include primary centering uncertainties, which dominate positional uncertainty here.}. Furthermore, besides redetecting all known background sources with similar astrometric and photometric values to the initial analysis, we imaged an additional faint source (CC14) owing to \verb+PACO+'s deeper sensitivity. Astrophotometrical results for B, C, Cb and CC14 are provided in Appendix \ref{Appendix_astro_photo}.

%=============================================
\section{A low-mass substellar companion around HIP~81208~C}
\label{sec:Cb_properties}
%=============================================
%
\subsection{Companion confirmation}

\begin{figure}[t!]
    \centering
    \includegraphics[width=\linewidth]{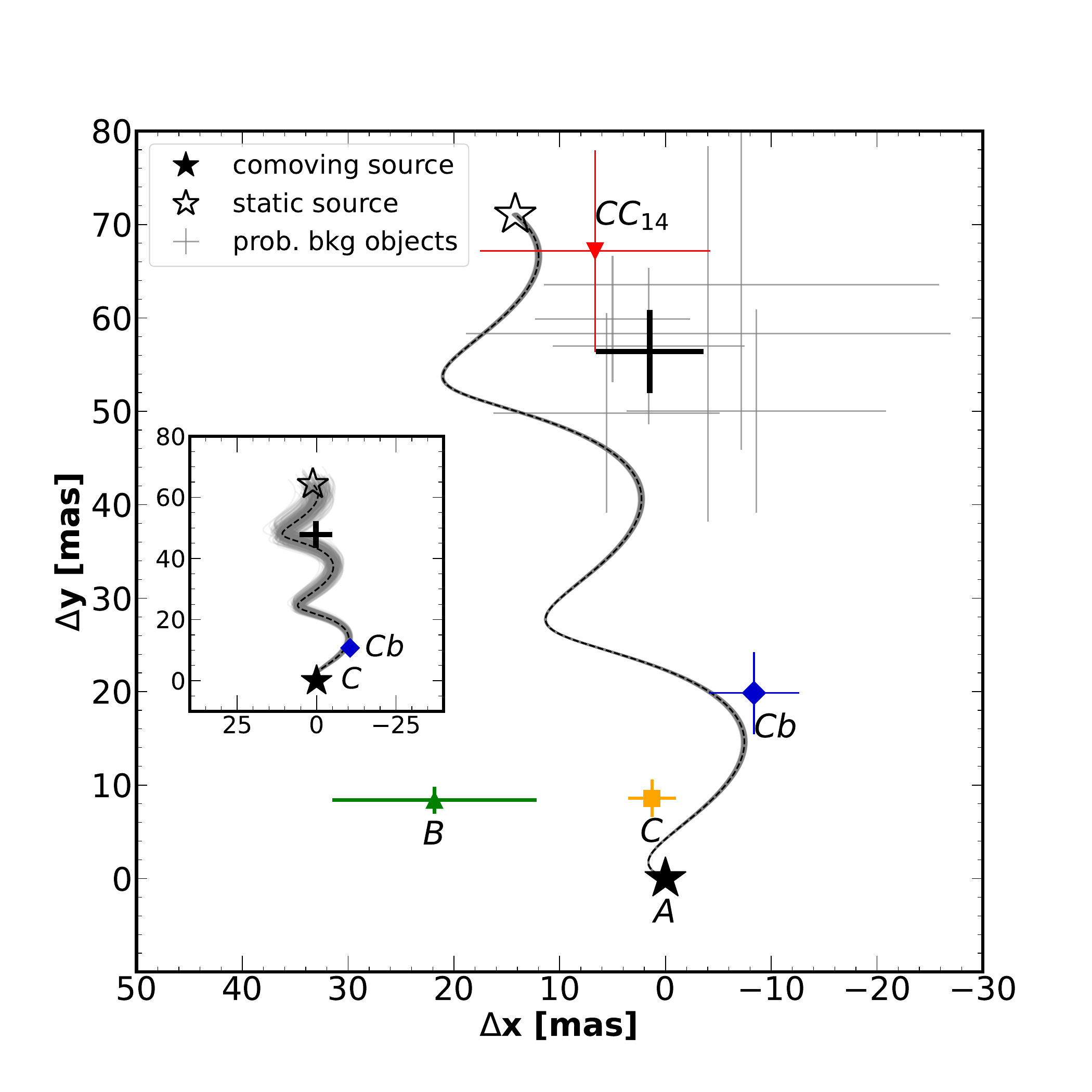}
    \caption{Astrometric displacements between the two epochs for all known sources \citep[data from][]{Viswanath_hip81208_tmp} with the addition of Cb (blue diamond) and CC14 (red down-pointing triangle). The expected displacement at the second epoch is indicated by a white and a black star for a static background source and a co-moving source with the target, respectively. The mean astrometric shift of background sources (gray crosses) is shown as a thick gray cross. Inset: relative motion between Cb and C, with C set as the primary; the error bar associated to Cb is small enough to be hidden behind the diamond.}
    \label{fig:astrometric_displacements}
\end{figure}

Figure \ref{fig:astrometric_displacements} shows the proper motion of B, C and Cb as opposed to already known background sources. As for B and C, the motion of Cb is inconsistent (at $6 \sigma$) with the observed motion of field interlopers. We anticipate that the non-null relative motion between C and Cb is consistent with orbital motion, as it will be quantified in Sect.~\ref{subsec:orbital_properties}.

Having confirmed the common proper motion of Cb to the already known HIP~81208~A, B and C, we carefully investigated an alternative possibility: namely, that A-B and C-Cb constitute two independent binary members of UCL projected at a short separation (Appendix~\ref{Appendix_FAP}). We therefore assessed the probability that an UCL member, unseen by Gaia \citep{gaia_dr3}, could end up as interloper to any BEAST star. A final false alarm probability, referring to at least one detection across the entire current survey (47 stars), of $1.3 \times 10^{-3}$ was obtained (meaning that we expect, on average, 1 false positive out of $\sim$40000 observed B stars), placing a high confidence level on the membership of A (+B) and C (+Cb) to a single quadruple system. 

\subsection{Physical properties}
\label{sec:Cb_astro_parameters}

As in \citet{Viswanath_hip81208_tmp}, an estimate of the photometric mass of the newly discovered object was obtained using the \textsc{madys} tool \citep{madys}: we combined \verb+PACO+ ($K1$, $K2$) contrasts, the 2MASS $K_s$ magnitude of the primary \citep{2MASS}, plus the system's color excess ($\ebv=0.011\pm0.021$ mag) and age ($17^{+3}_{-4}$~Myr). For the purpose of accounting for theoretical uncertainties on the final estimates, the computation was performed by comparison with two different models suited for the age and mass range of interest: the Ames-Dusty models \citep{ames_models} and the BT-Settl models \citep{bt_settl_models}. The resulting values were averaged to yield a final mass estimate:

\begin{equation}\label{eq:Cb_mass}
M_{\text{Cb}} = 14.8 \pm 0.4~\Mjup 
\end{equation}

Additional details on the derivation of this estimate and its associated uncertainty are provided in Appendix~\ref{Appendix_Cb_mass}. Based on models, this best-fit mass would correspond to expected 2MASS $H=10.28\pm0.07$ mag and $K_s=9.74 \pm 0.04$ mag.

In the same fashion, the average effective temperature, surface gravity and bolometric luminosity returned by this comparison are $\teff = 2050_{-20}^{+35}$ K, $\log{g}=4.087^{+0.011}_{-0.022}$, and $\log{L/\Lsun}=-3.31 \pm 0.03$. 

However, we acknowledge that we currently have only 2 photometric points probing similar wavelengths and that, even in the case of young substellar objects  with much more comprehensive data available, systematic errors intrinsic to atmosphere and evolution models might be up to an order of magnitude larger than formal uncertainties \citep[see, e.g.,][]{theo_unc}. These differences can arise for instance from uncertainties on the initial entropy after accretion, possible age difference between planet and host star as well theoretical difficulties in handling atmospheric dust \citep[see, e.g.,][]{2023arXiv230507719L}.

\begin{figure}[t!]
    \centering
    \includegraphics[width=0.8\linewidth]{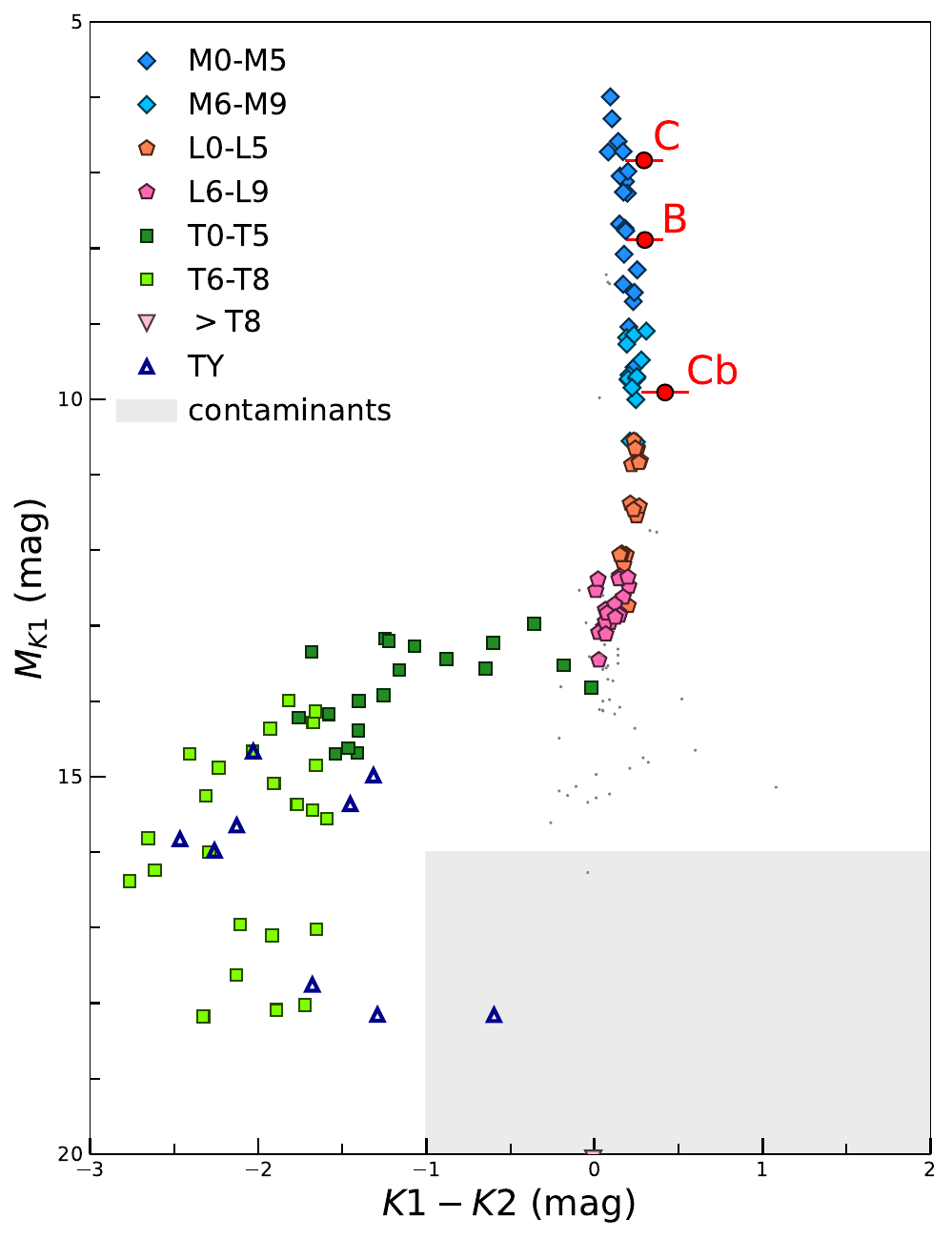}
    \caption{Color magnitude diagram with the 3 components (B, C and Cb) represented by the red dots.}
    \label{fig:CMD}
\end{figure}

Figure \ref{fig:CMD} displays the position of B, C and Cb on a color-magnitude diagram, all matching the M sequence, while Table~\ref{tab:recap_table} reports the main outputs of the astrophotometric characterization of the object.

\begin{figure}[t!]
    \centering
    \includegraphics[width=\linewidth]{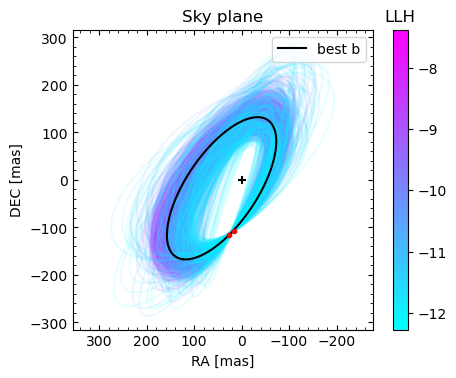}
    \caption{Motion of Cb around C in RA / DEC for the 1000 best draws from our posterior distributions. LLH stands for log likelihood.}
    \label{fig:orbital_fit}
\end{figure}

\begin{table*}
\caption{Final astrometry for Cb with respect to A (sep, PA) and C ($\Delta$RA, $\Delta$DEC), measured IRDIS contrasts and best-fit values for mass and $\teff$. The mass ratio with respect to C is indicated as $q_{Cb}$.}
\label{tab:recap_table}
\centering
\begin{tabular}{lcccccc|ccc}
\hline\hline
epoch & sep (mas) & PA ($^\circ$) & $\Delta K_1$ (mag) & $\Delta K_2$ (mag) & $\Delta$RA (mas) & $\Delta$DEC (mas) & mass ($\mjup$) & $q_{Cb}$ & $\teff$ ($K$) \\
%& (mas) & ($^\circ$) & (mag) & (mag) & (mas) & (mas) & ($\mjup$) & & (K) \\
\hline 
2019 & $1573.4 \pm 3.2$ & $119.9 \pm 0.12$ & $8.93 \pm 0.08$ & $8.59 \pm 0.08$ & $26.4 \pm 1.13$ & $-116.9 \pm 0.99$ & \multirow{2}{*}{$14.8 \pm 0.4$} & \multirow{2}{*}{$
0.11 \pm 0.01$} & \multirow{2}{*}{$2050^{+35}_{-20}$} \\
2022 & $1556.3 \pm 2.8$ & $119.42 \pm 0.11$ & $8.99 \pm 0.12$ & $8.49 \pm 0.13$ & $15.6 \pm 1.64$ & $-107.4 \pm 0.47$ & & & \\
\hline
\end{tabular}
\end{table*}

\subsection{Orbital properties}
\label{subsec:orbital_properties}
We ran the \verb+emcee+ code \citep{2013PASP..125..306F} to derive information on the orbital parameters of the companion starting from the astrometry and their best-fit masses of Cb and C. We used their relative astrometry, as measured by PSF subtraction, because it is affected by much smaller uncertainties than absolute astrometry (Appendix~\ref{Appendix_PSF}). The sampling tool is based on the \verb+emcee+ 3.0 library, using a mix of custom move functions to alleviate potential multi-modality problems and the cyclicity of angular variables. 100 walkers, between 100000 and 400000 iterations, and 10 temperatures were used. The priors include a uniform log  prior for sma ($a \in [0,80]$ au). The upper value in sma corresponds roughly to  0.3 times the projected separation between A and C, which, following \cite{stability} 
should allow for the stability of Cb in the binary system, given the masses of A and C, and assuming a null eccentricity. We nonetheless considered eccentricities $e \in [0,0.4]$ range for the priors\footnote{Note that assuming a larger range (up to 0.9) does not significantly change the results.}. 

Given the limited information available, the orbital parameters are poorly constrained (see Fig. \ref{fig:corner_plot_Cb} and in  Appendix~\ref{Appendix_corner_plot}). The $a$ distribution peaks at 17 au ($T \sim 190$ yr), with a tail extending to more than 40 au ($T \sim 600$ yr). The eccentricity is not constrained. The inclination of its orbital plane is  $i=73\pm 20^\circ$. Figure \ref{fig:orbital_fit} shows the 1000 best draws from our posterior distributions for Cb determined based on the loglikelihood. We also ran the MCMC on B and C companions, finding $a$, $e$ and $i$ compatible with those found by \citet{Viswanath_hip81208_tmp}, albeit with larger uncertainties, due to the larger error bars found in the present astrometric measurements. 

\section{Discussion and concluding remarks}
\label{sec:conclusion}
%inclure additional planets ?

Before BEAST began, no $\lesssim 30 \mjup$ companion was known around stars more massive than $3 \Msun$ -- with only sporadic detections by RV in the mass range $2.5-3 \Msun$ \citep{ny_oph,wolthoff22}, questioning their very existence \citep[see discussion in][]{Janson_beast}. The discovery of a circumstellar $\sim 11~\Mjup$ planet around the $6-10~\msun$ binary $b$ Centauri \citep{b_centauri} and one (possibly two) brown dwarfs close to the deuterium-burning limit around the $\sim 9~\msun$ $\mu^2$ Scorpii \citep{mu2_sco} first provided evidence for such a population, opening up a plethora of questions about its genesis.

The architecture of the HIP~81208 system turns out to be unique in many respects (Table~\ref{tab:system_architecture}). Not only is the B-type primary surrounded by a brown dwarf and a M-type stellar companion; the additional discovery, presented in this Letter, of a $\sim 15 \mjup$ companion to the C component makes it the first binary system with substellar companions to both components ever discovered by imaging.

\begin{table}[t!]
    \centering
    \begin{tabular}{lcccc}
            Body & Primary & M ($M_\odot$) & $a$ (au) & $T$ (yr) \\
            \hline
            \hline
        A & A & $2.58 \pm 0.06$ & \textemdash & \textemdash \\
        B & A & $0.064^{+0.006}_{-0.007}$ & $53.98^{+32.22}_{-15.00}$ & $246.9^{+251.3}_{-95.4}$ \\
        C & A & $0.135^{+0.010}_{-0.013}$ & $234.27^{+168.65}_{-68.96}$ & $2232.4^{4429.4}_{-1213.6}$ \\
        Cb & C & $0.0141 \pm 0.0004$ & $23.04^{+13.88}_{-6.55}$ ${(a)}$  & $285.00^{+293.67}_{-112.07}$ \\
        \hline
        \hline
    \end{tabular}
    \caption{Emerging architecture of the hierarchical HIP 81208 system, with median and ($16^{th}$,$84^{th}$) percentiles used for $a$ and $T$. }
    \label{tab:system_architecture}
    \vspace{0.2cm}
    \RaggedRight
    \textbf{Notes:} Values taken from \cite{Viswanath_hip81208_tmp} for B and C. ${(a)}$ Note that the distribution of the semi-major axis of Cb is not gaussian and the peak of the distribution is at about 17 au (see text). The column "primary" indicates the parent body of each object, to which $a$ and $T$ also refer.
\end{table}

\begin{figure}[t!]
    \centering
    \includegraphics[width=\linewidth]{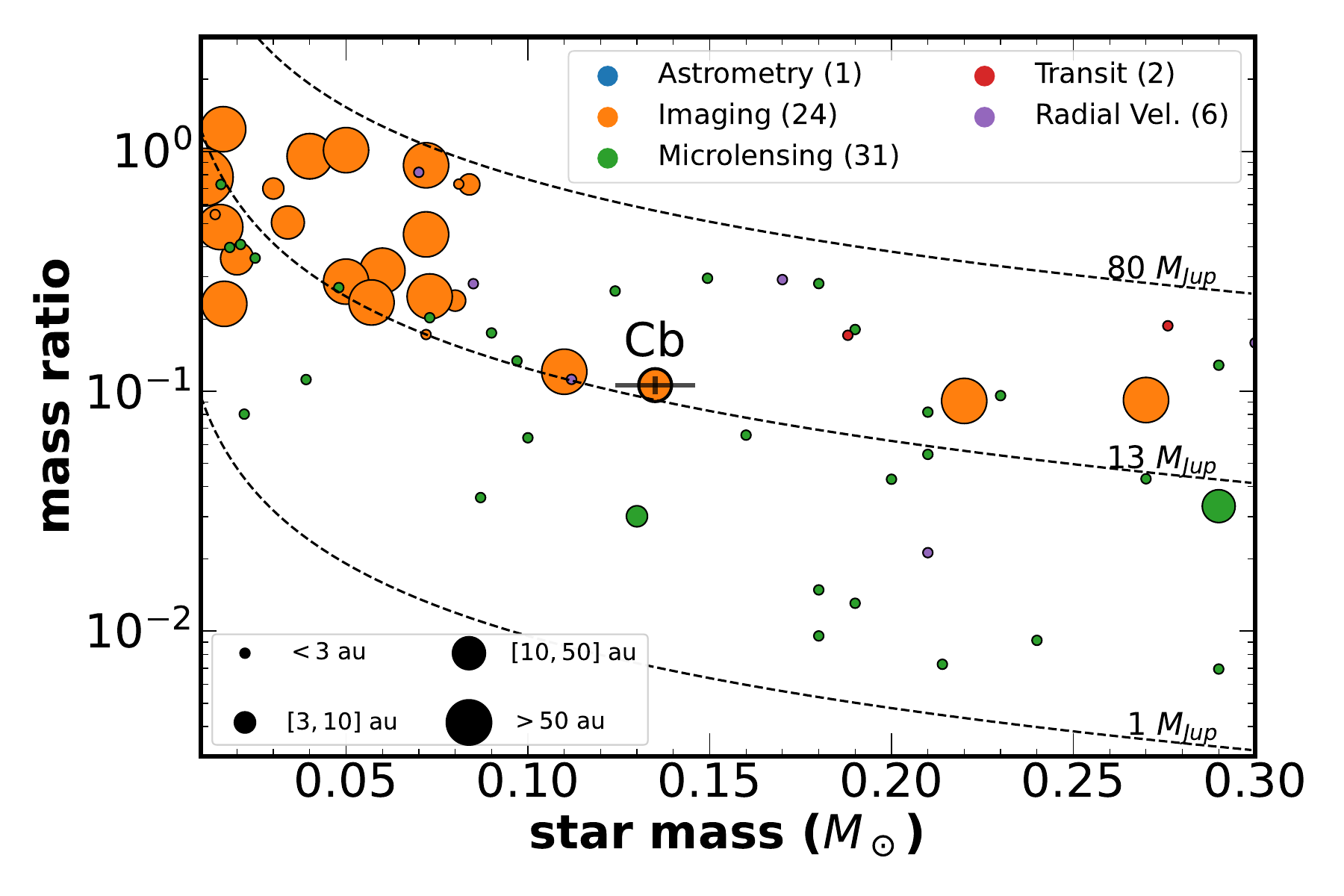}
    \caption{Mass ratio vs stellar mass for confirmed giant planets and brown dwarf companions ($1~\mjup<M<80~\mjup$) to late M-type and massive brown dwarf primaries ($0.03~\msun<M<0.3~\msun$). Masses of RV companions actually correspond to $M \sin{i}$. Detection methods are labeled through different colors; semimajor axes or, if not available, projected separations have been grouped in four bins, each one associated to a circle size. The position of HIP~81208, provided with errorbars, is explicitly labeled.}
    \label{fig:q_plot}
\end{figure}

Even if considered in isolation, a $\simeq 15~\Mjup$ companion at $\sim 20$ au from a late M-type star such as HIP~81208~C would be deemed remarkable. Figure~\ref{fig:q_plot} shows the mass ratio of confirmed giant planets and brown dwarfs ($M \in [1, 80]~\Mjup$) around late M-type stars ($M_* \in [0.08,0.3]~\Msun$): among five such DI companions, only two -- WISE J0720-0846 \citep{mstar_companions1} and LHS 2397a B \citep{mstar_companions2} -- have orbital separations $<50$ au; both of them, however, are characterized by a much larger mass ratio ($q \approx 0.7$) than $q_{Cb}$, indicative of a binary-like formation \citep{binary_like}. Including indirect techniques (with $a \in [3,50]$ au), only two $a<10$ au objects, discovered via microlensing -- OGLE-2016-BLG-0263L b \citep{mstar_companions3} and OGLE-2013-BLG-0911L b	\citep{mstar_companions4}, both with a small $q \approx 0.03$ -- are added\footnote{Data from the Exoplanet Encyclopaedia (\href{http://exoplanet.eu/}{http://exoplanet.eu/}).}. 
%Bridging the gap between companions discovered by direct and indirect techniques, HIP~81208~Cb stands out as a particularly appealing target for follow-up studies bereft of differential instrument systematics.

While a full formation analysis of HIP~81208~Cb is beyond the scope of this work, it is worth mentioning possible formation pathways for the object and the whole quadruple system. Pivotal to a full understanding of the observed architecture is the formation of HIP 81028 C: the M-star could be an outcome of either turbulent fragmentation of a star-forming core \citep{fragmentation1} -- possibly followed by inward migration \citep{kuffmeier19} -- or gravitational instability (GI) within the disk of HIP~81208~A \citep{GI_stars}, the rough cutoff between these mechanisms ($\sim 500$ au) likely depending on environment and stellar mass \citep{binary_mechanism_cutoff}.

If the former is true, the circumstellar disks of A and C would be truncated due to mutual gravitational actions \citep{panic21}. Provided no significant alteration of initial orbital parameters, a tentative estimate of the truncation radii $R_{T}$ for the two stars could be derived as in \citet{truncation_disk} by drawing $10^6$ ($a$,$e$) values for the A-C pair from the corresponding posterior distributions:
\begin{equation}
R_{T,C} = 0.88 \cdot R_{R} = 0.88 \cdot r_p \cdot \frac{0.49 \cdot q^{2/3}}{0.6 \cdot q^{2/3} + \ln{(1+q^{1/3})}} = 26_{-9}^{+16}~\text{au},
\end{equation}
and 
\begin{equation}
R_{T,A} = 0.88 \cdot (r_p-R_{R}) = 130_{-40}^{+70}~\text{au},
\end{equation}

where $r_p$ indicates the periastron of the orbit, and the Roche lobe $R_{R}$ is derived from the empirical formula by \citet{roche_lobe}. The current position of Cb would be only marginally within the locations of its alleged parent disk, whence it would have sprouted either via CA \citep{core_accretion,alibert13,emsenhuber21} or GI. According to CA models, the formation of a high-$q$ $15 \mjup$ object around a late M-type star is not expected \citep{2008ApJ...673..502K,adams21,schlecker22}, also due to formation timescales ($\sim 10^{6-7}$ Myr) exceeding typical disk lifetimes by one or two order of magnitude at separations $\gtrsim 10$ au  \citep{2009ApJ...707...79D}. Conversely, GI could represent a formation channel for Cb \citep{GI_paper3,GI_paper4,GI_paper2}, provided a unusually large disk-to-star ratio of $\sim 30\%$ ($\approx 40 \mjup$) compared to the expected 1-10\% \citep[][and references therein]{2018A&A...618L...3M,haworth20,Mercer_2020}. Interestingly enough, the C-Cb separation would be within the typical range of M-type stellar binaries \citep{winters19,susemiehl22}.

According to the alternative scenario, C itself would have formed via GI within the disk of A: simulations are able to produce objects with masses as high as $0.12 \Msun$ already around $<1.2 \Msun$ stars \citep{GI_paper6}; despite the lack of GI models suited to B-type hosts, evidence for a more-than-linear dependence of disk mass on stellar mass \citep{disk_mass_ratio}, coupled to observed circumstellar disks to late B stars spanning hundreds of au \citep{disk_size1,disk_size2,disk_size3}, tentatively hints towards such possibility. As a companion to a disk-borne object, HIP~81208~Cb would intriguingly retain -- whether formed in-situ or via dynamical capture \citep{podsiadlowski10,ochiai14} -- the hierarchical level of a satellite \citep{lazzoni22}.

A detailed characterization of the orbital parameters of B, C and Cb, and in particular their mutual inclinations, will discriminate between the two scenarios, helping shed light on this unique multiple system.

\begin{acknowledgements}
This project has received funding from the European Research Council (ERC) under the European Union's Horizon 2020 research and innovation programme (COBREX; grant agreement n° 885593). 
SPHERE is an instrument designed and built by a consortium consisting of IPAG (Grenoble, France), MPIA (Heidelberg, Germany), LAM (Marseille, France), LESIA (Paris, France), Laboratoire Lagrange (Nice, France), INAF - Osservatorio di Padova (Italy), Observatoire de Genève (Switzerland), ETH Zürich (Switzerland), NOVA (Netherlands), ONERA (France) and ASTRON (Netherlands) in collaboration with ESO. SPHERE was funded by ESO, with additional contributions from CNRS (France), MPIA (Germany), INAF (Italy), FINES (Switzerland) and NOVA (Netherlands). SPHERE also received funding from the European Commission Sixth and Seventh Framework Programmes as part of the Optical Infrared Coordination Network for Astronomy (OPTICON) under grant number RII3-Ct-2004-001566 for FP6 (2004-2008), grant number 226604 for FP7 (2009-2012) and grant number 312430 for FP7 (2013-2016). 
This work has made use of the SPHERE Data Centre, jointly operated by OSUG/IPAG (Grenoble), PYTHEAS/LAM/CeSAM (Marseille), OCA/Lagrange (Nice), Observatoire de Paris/LESIA (Paris), and Observatoire de Lyon (OSUL/CRAL).
T.H. acknowledges support from the European Research Council under the Horizon 2020 Framework Program via the ERC Advanced Grant Origins 83 24 28.
G-DM acknowledges the support of the DFG priority program SPP 1992 Exploring the Diversity of Extrasolar Planets (MA~9185/1) and from the Swiss National Science Foundation under grant 200021\_204847 PlanetsInTime. Parts of this work have been carried out within the framework of the NCCR PlanetS supported by the Swiss National Science Foundation.
R.G. and S.D. acknowledge the support of PRIN-INAF 2019 Planetary Systems At Early Ages (PLATEA). 
This research has made use of the SIMBAD database and VizieR catalogue access tool, operated at CDS, Strasbourg, France. This work is supported by the French National Research Agency in the framework of the Investissements d’Avenir program (ANR-15-IDEX-02), through the funding of the "Origin of Life" project of the Univ. Grenoble-Alpes.
This work was supported by the Action Spécifique Haute Résolution Angulaire (ASHRA) of CNRS/INSU co-funded by CNES.
This research has made use of data obtained from or tools provided by the portal exoplanet.eu of The Extrasolar Planets Encyclopaedia.
\end{acknowledgements}

\bibliographystyle{aa}
\bibliography{sample.bib}
%TC:ignore
\newpage
\begin{appendix}
\section{Details on the custom PSF subtraction routine}
\label{Appendix_PSF}

The custom PSF subtraction routine, sketched in Sect~\ref{sec:data_reduction}, proceeded in two steps based on small stamps of 47$\times$47 pixels roughly centered on C. In the first step we used as a model a 47x47 pixel stamp of the mean of the two off-axis PSF, acquired just before and just after the observations. We recursively removed C and then Cb from each of the 48 individual frames of the coronagraphic sequence by minimizing the residuals between the data stamp and our empirical PSF model stamp. This minimisation was performed by injecting negative versions of this empirical PSF model in a local grid centered on C  oversampled by a factor 100 in each spatial dimension. We used three free parameters each for C and Cb, namely the oversampled pixels positions in x and y and the source to model contrast, and selected the models that minimise the absolute value of the residuals on small optimisation zones (1000 by 1000 oversampled pixels).
After this first step, we noticed that the residuals after subtraction of this first PSF model were characterized by 1) a relatively high intensity, of the order of 1-2\% of the local flux of C and 2) a systematic shape independent of time but with an alignment following the parallactic angle rotation. We interpreted these features as hints that the local PSF at C's position slightly differed from the calibration off-axis PSF. We therefore added C again (using its fitted parameters) on each of the 48 stamps with both C and Cb removed, effectively building a local PSF with the same pupil orientation on each frame. Afterwards, we applied a similar approach to standard ADI, which median-combined the 48 resulting sub-frames obtained at different parallactic angles without derotating them, producing a local pupil-stabilised PSF. As in ADI, the weak residuals from the subtraction of Cb, already close to the background noise and rotating around C with the parallactic angle, were further removed from this local PSF model by means of the non-derotated median. 

As a second step, we repeated the same minimisation approach, starting from the 48 small stamps containing both C and Cb, but using this local pupil-stabilised PSF as model instead of the off-axis PSF. Minimization directly provides the best-fit parameters for position and contrast for C and Cb on each frame, albeit in a pupil-tracking rotating frame of reference. These measurements were then derotated to sky coordinates and averaged to obtain the results shown in Table \ref{tab:astro_photo_cc}. The intensity of the residuals was reduced by a factor $\sim$5 after this second step compared to the first step, the absolute value of the highest residuals in any individual frames at both epochs being barely above the local background noise.

We determined the uncertainties associated with our estimates of positions and flux by deriving the standard deviation of the resulting 48 independent measurements, which naturally and robustly include all sources of systematic errors that cause frame to frame variations, such as tip-tilt jitter or atmospheric transmission variability during the observing sequence. The main remaining systematic -- namely the uncertainty on the position of the central star, constant over the sequence -- was quadratically added to the measured uncertainties and dominates the error budget. In the peculiar case of the relative position of Cb around C (reported in Table~\ref{tab:recap_table}), this systematic is naturally canceled out and the dynamical fits performed in subsection~\ref{subsec:orbital_properties} consequently employ the much smaller errors bars obtained when this systematic contribution is taken out.

\begin{figure}[t!]
    \centering
    \includegraphics[width=\linewidth]{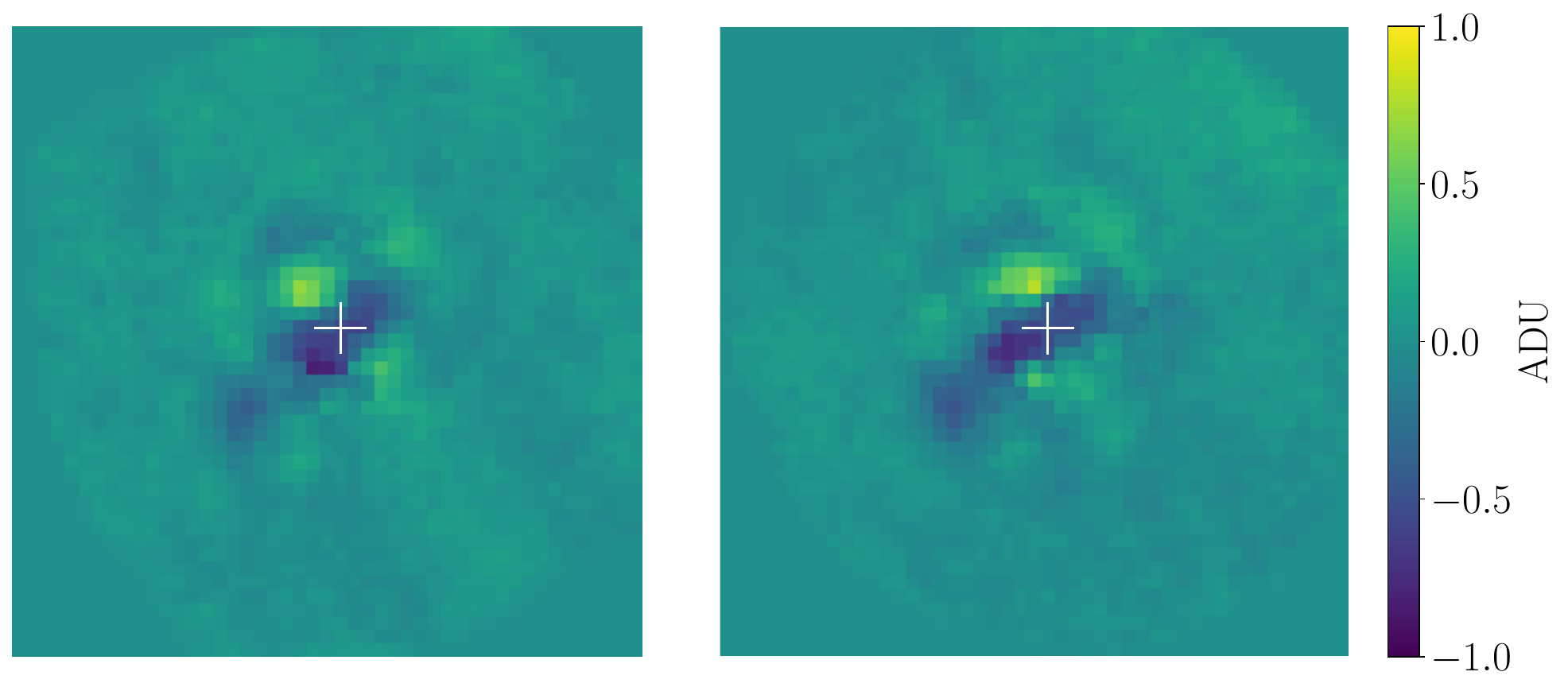}
    \caption{Residuals after subtracting both the PSF of C and Cb (2019 on the left, 2022 on the right) for the K1 band. The position of C (subtracted) is highlight by the white cross. Residuals are below 1 ADU.}
    \label{fig:residual_custom_psf_sub}
\end{figure}

\section{Raw coronagraphic frames displaying the ADI rotation around C of Cb}
\label{Appendix_raw_frames}

\begin{figure*}[t!]
    \centering
    \begin{subfigure}[b]{0.33\textwidth}
        \centering
        \includegraphics[width=\textwidth]{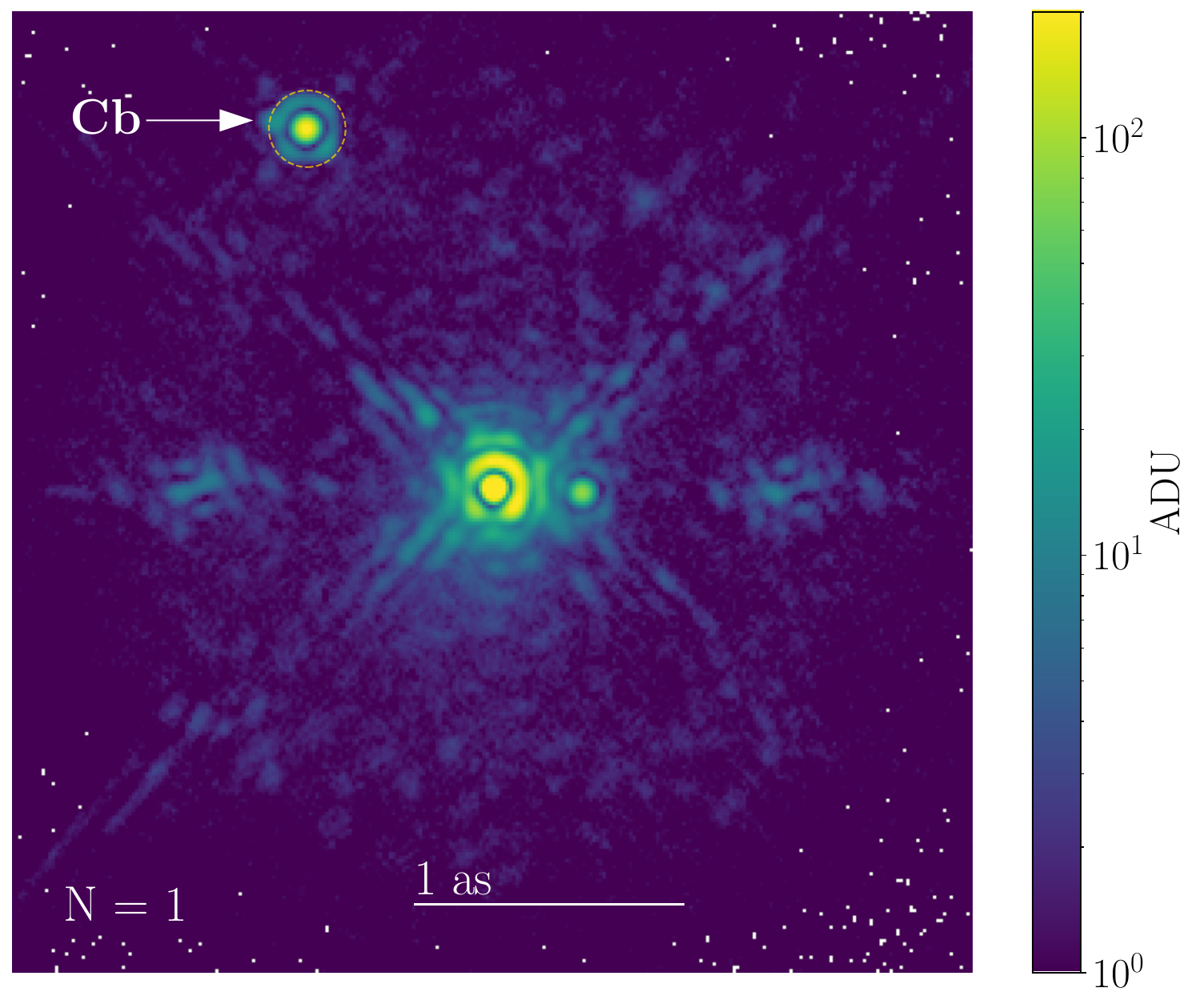}
        \caption[]
        {{\small First raw coronagraphic frame of the 2019 dataset}}    
        \label{subfig:frame_1}
    \end{subfigure}
    \hfill
    \begin{subfigure}[b]{0.33\textwidth}  
        \centering 
        \includegraphics[width=\textwidth]{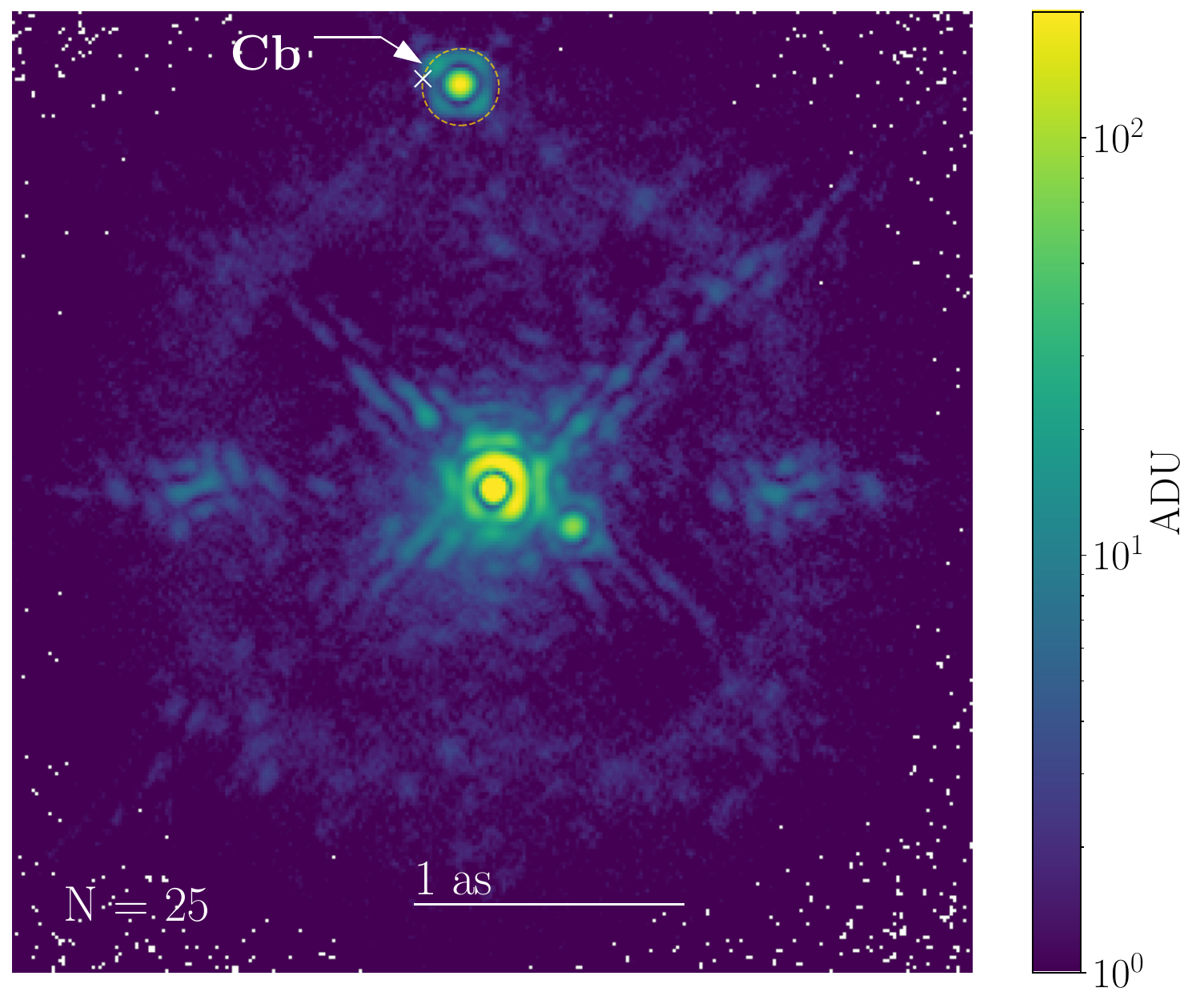}
        \caption[]%
        {{\small 25th raw coronagraphic frame of the 2019 dataset}}    
        \label{subfig:frame_24}
    \end{subfigure}
    \hfill
    \begin{subfigure}[b]{0.33\textwidth}  
        \centering 
        \includegraphics[width=\textwidth]{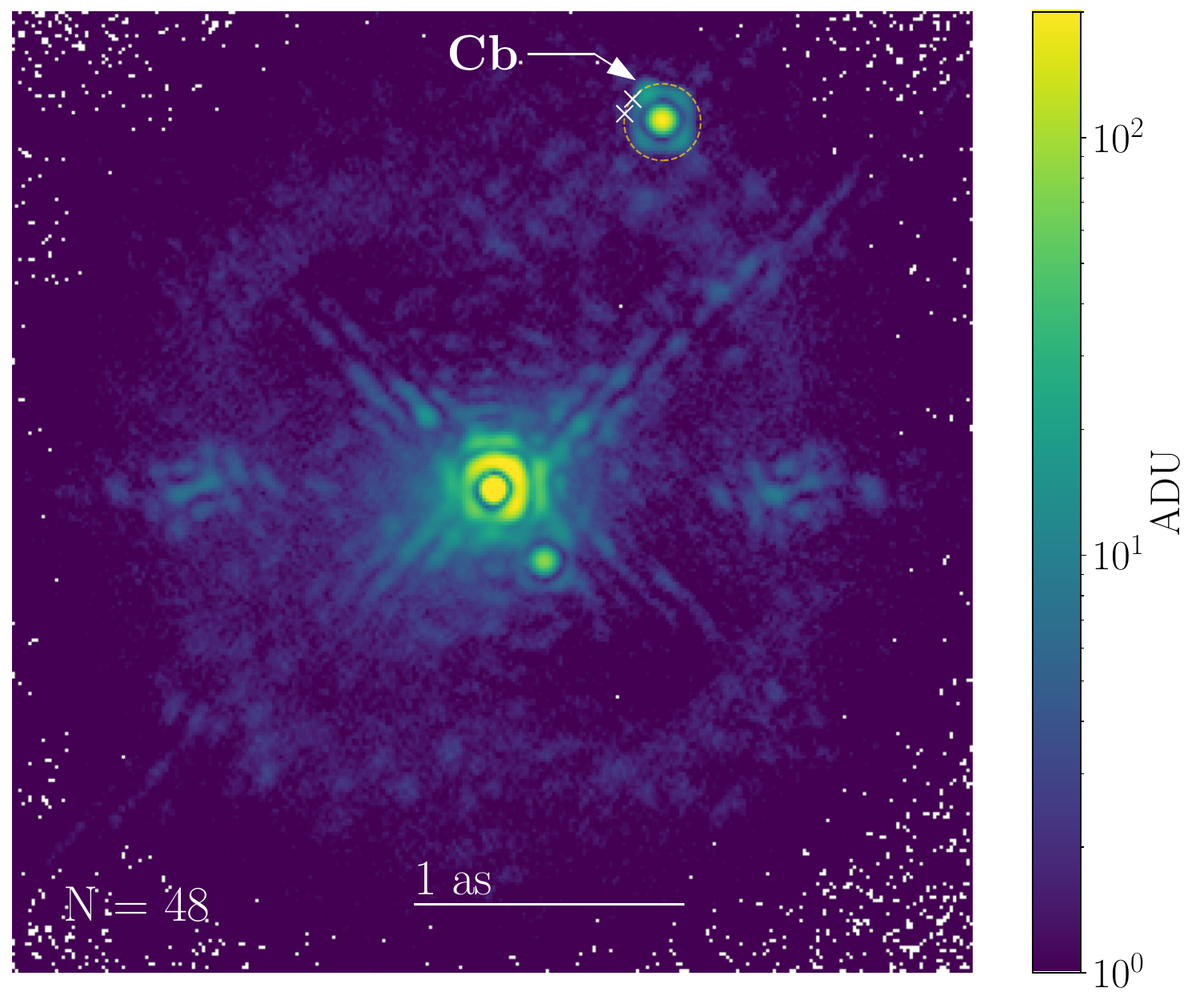}
        \caption[]%
        {{\small Last raw coronagraphic frame of the 2019 dataset}}    
        \label{subfig:frame_48}
    \end{subfigure}
    \caption[]
    {Collection of raw coronagraphic frames of the 2019 dataset. The current position of the Cb companion at the given frame is highlighted by the white arrow and past positions (projected onto the current frame) are highlighted by the white crosses. We can clearly see the expected rotation due to the ADI sequence. A PSF artifact would have resided at a fixed angle relative to C in all frames.} 
    \label{fig:frame_sequence}
\end{figure*}

As mentioned in Sect.~\ref{subsec:det_sources}, the presence of the source Cb is visually evident even in raw coronagraphic frames despite its close vicinity to HIP 82108 C. A subset of the 48 individual frames is provided in Fig.~\ref{fig:frame_sequence}.

\section{Astrophotometric results}
\label{Appendix_astro_photo}

We provide in Table~\ref{tab:astro_photo_cc} a summary of the astrometric and photometric results obtained for HIP~81208~Cb (Sec.~\ref{subsec:det_sources}) by means of the PACO and PSF subtraction reductions described in Sec.~\ref{sec:data_reduction}. Values for the newly found background source CC14 -- only detected by \verb+PACO+ -- are provided in Table~\ref{tab:bckg_additional}.

\begin{table*}[t!]
\tiny
\centering
%\begin{adjustbox}{width=\linewidth,center}
    % \begin{adjustbox}{center}
        \begin{tabular}{ll|lll|lll|}
            & & \multicolumn{3}{c}{2019-08-06} & \multicolumn{3}{|c|}{2022-04-05} \\\cline{3-8}
            & & HIP~81208 B & HIP~81208 C & HIP~81208 Cb & HIP~81208 B & HIP~81208 C & HIP~81208 Cb\\
            \hline
           % \multirow{4}{*}{No ADI} & \multicolumn{1}{l|}{Sep (mas)} & 320.6$\pm$1.4 & 1494.6$\pm$2.3 & 1580.3$\pm$2.6 & 328.5$\pm$0.8 & 1491.1$\pm$2.0 & 1579.3$\pm$2.8\\\cline{2-8}
           %                      & \multicolumn{1}{l|}{PA ($\degree$)} & 358.94$\pm$7.6 & 116.23$\pm$0.07 & 119.88$\pm$0.07 & 0.44$\pm$0.13 & 115.97$\pm$0.05 & 119.65$\pm$0.06\\\cline{2-8}
           %                      & \multicolumn{1}{l|}{$\Delta$K1 (mag)} & 7.05$\pm$0.09  & 5.80$\pm$0.06 & 9.17$\pm$0.14 & 6.84$\pm$0.12 & 5.76$\pm$0.12 & 10.35$\pm$0.97 \\\cline{2-8}
            %                     & \multicolumn{1}{l|}{$\Delta$K2 (mag)} & 6.76$\pm$0.08  & 5.55$\pm$0.07 & 9.52$\pm$0.12 & 6.58$\pm$0.12 & 5.50$\pm$0.12 & 9.50$\pm$0.57 \\ \cline{1-8}
            \multirow{4}{*}{PACO ASDI } & \multicolumn{1}{l|}{Sep (mas)} & 315.5$\pm$3.1 & 1495.1$\pm$3.2 & 1573.4$\pm$3.2 & 324.4$\pm$2.7 & 1493.0$\pm$2.8 & 1556.3$\pm$2.8\\\cline{2-8}
                                 & \multicolumn{1}{l|}{PA ($\degree$)} & 357.6$\pm$0.57  & 116.3$\pm$0.13 & 119.9$\pm$0.12 & 1.39$\pm$0.48 & 115.95$\pm$0.09 & 119.42$\pm$0.11 \\\cline{2-8}
                                 & \multicolumn{1}{l|}{$\Delta$K1 (mag)} & 6.99$\pm$0.07  & 5.85$\pm$0.08 & 8.93$\pm$0.08 & 6.91$\pm$0.11 & 5.93$\pm$0.12 & 8.99$\pm$0.12 \\\cline{2-8}
                                 & \multicolumn{1}{l|}{$\Delta$K2 (mag)} & 6.70$\pm$0.08  & 5.48$\pm$0.08 & 8.59$\pm$0.08 & 6.60$\pm$0.16 & 5.49$\pm$0.12 & 8.49$\pm$0.13 \\ \cline{1-8}
            \multirow{4}{*}{PSF subtraction} & \multicolumn{1}{l|}{Sep (mas)} & - & 1494.4$\pm$3.2 & 1573.1$\pm$3.3 & - & 1491.4$\pm$2.8 & 1554.8$\pm$3.2\\\cline{2-8}
                                 & \multicolumn{1}{l|}{PA ($\degree$)} & -  & 116.36$\pm$0.14 & 119.76$\pm$0.15 & - & 115.89$\pm$0.15 & 119.2$\pm$0.16 \\\cline{2-8}
                                 & \multicolumn{1}{l|}{$\Delta$K1 (mag)} & - & 5.80$\pm$0.03 & 8.96$\pm$0.04 & - & 5.81$\pm$0.03 & 8.97$\pm$0.08 \\\cline{2-8}
                                 & \multicolumn{1}{l|}{$\Delta$K2 (mag)} & - & 5.54$\pm$0.03 & 8.55$\pm$0.04 &  - & 5.48$\pm$0.03& 8.45$\pm$0.08 \\
          \hline
        \end{tabular}
    %\end{adjustbox}
%     \vspace{ - 05 mm}
    \caption{Astrometry and photometry extracted for the companions B, C and Cb the two algorithms. The astrometric values are averaged over the two wavelengths.}
    \label{tab:astro_photo_cc}
\end{table*}

\begin{table}[t!]
\tiny
    \centering
    \begin{tabular}{c|cccc}
         & Sep (mas) & PA ($\degree$) & $\Delta$K1 & $\Delta$K2 \\
         \hline
    2019-08-06 & 3923.3$\pm$6.7 & 141.1$\pm$0.1 & 14.57$\pm$0.26 & 14.03$\pm$0.27\\
    2022-04-05 & 3875.5$\pm$8.14 & 140.4$\pm$0.13 & 15.87$\pm$0.28 & 15.04$\pm$0.26 \\
    \hline
    \end{tabular}
    \caption{Astrometry and photometry for the additional background source CC14 detected in this analysis.}
    \label{tab:bckg_additional}
\end{table}

\section{On the bound nature of A-B and C-Cb}
\label{Appendix_FAP}

Whilst the proper motion analysis of Cb firmly allowed us to conclude on its non-background nature and its common motion to A, B and C, it does not excluded, in principle, an alternative hypothesis:
\begin{quote}
    The A-B system is totally independent of the C-Cb system, both being Sco-Cen binaries projected by chance at a short separation from one another.
\end{quote}
In order to quantify the probability of this alternative scenario, we adapted the argument already produced for HIP~81208~B \citep{Viswanath_hip81208_tmp} and $\mu^2$ Scorpii b \citep{mu2_sco}.
After defining indicative coordinate limits for UCL as $(l, b) = [313^\circ,343^\circ] \times [2^\circ,28^\circ]$, we recovered $N_0=3835$ bona-fide members to this subgroup from the Gaia DR2-based list of Sco-Cen members assembled by \citet{damiani_scocen}. At the distance and age of Sco-Cen, the census of the stellar population of the association is reasonably complete\footnote{According to BHAC15 isochrones \citep{bhac15_models} at solar metallicity, a 0.08 \msun~star aged 15 Myr has absolute $G=11.45$, corresponding to an apparent $G \sim 17$ mag at the mean separation of UCL ($\sim 140$ pc); the survey is virtually complete for $G \in [12,17]$ mag \citep{2018A&A...616A...1G}.}. However, a source can be overlooked by Gaia if it happens to be located too close to a brighter star, that is, if the $\Delta G$ between the former and the latter is larger than the maximum contrast achievable by the satellite at the corresponding angular separation $s$. Let us therefore define as {\textit{shaded area}} the circular region, centered on a star, within which the average detection efficiency $\bar{\delta}(s,\Delta G)$ of Gaia equals 50\% for a given apparent G magnitude, and effective separation $s_{\text{eff}}$ the corresponding radius.

Our goal here is to quantify the number of these \textit{phantom} UCL stars, so as to enable an estimation of the probability of spotting at least one of them within the entire BEAST survey. Intuitively, the computation hinges upon 1) the total shaded area $A_{s}$, obtained as the sum of individual shaded areas for all Gaia sources within the boundaries of UCL, and 2) the number -- corrected for completeness -- of UCL members, $N_{\text{UCL}}$. As regards the former, we queried Gaia DR3 \citep{gaia_dr3} finding approximately $8 \cdot 10^7$ stars within the coordinate limits of UCL. We then recovered from \citet{gaia_contrasts} the detection efficiency of Gaia DR2 as a function of $\Delta G$ and $s$, $\delta(s,\Delta G)$. In this way, we were able to compute, for every Gaia source $i$, the effective separation $s_{\text{eff}}$ as a function of the apparent G magnitude of a hypothetical phantom star, $G$:
\begin{equation}
s_{\text{eff},i} = s~|~\bar{\delta}(s,\Delta G_i) = 0.5,
\end{equation}
where 
\begin{equation}
\bar{\delta}(s,\Delta G_i) = \frac{1}{s^2} \int_{0}^{s} \delta(\tilde{s},\Delta G_i) \tilde{s}^2 \,d\tilde{s}
\end{equation}
and $\Delta G_i(G) = G-G_i$. Summation over Gaia stars yields the total shaded area as a function of $G$:
\begin{equation}
A_s(G) = \sum_i{\pi s_{\text{eff},i}^2}.
\end{equation}
The probability density function (PDF) of phantom stars can be now expressed as:
\begin{equation}
    n_P(G) = \frac{A_s(G)}{A_{\text{UCL}}-A_s(G)} \cdot N_{\text{UCL}} \cdot 0.5 \cdot \zeta_{\text{Gaia}}(G),
\end{equation}
where $\zeta_{\text{Gaia}}(G)$ is the apparent G magnitude PDF of UCL stars, and the factor 0.5 accounts for the expectation that 50\% of these object were already detected by Gaia.

To cope with the incompleteness of the initial mass function (IMF) of the UCL sample at its faint end (i.e., for unseen substellar objects), we recovered the sample of Upper Scorpius\footnote{Upper Scorpius is, together with UCL and Lower Centaurus-Crux, one of the three subgroups in which Sco-Cen is classically divided \citep{scocen_subgroups}. We verified through a Kolmogorov-Smirnov test with $\alpha=0.05$ that the absolute G magnitude distribution for US stars -- selected from the same sample adopting coordinate boundaries as in \citet{usco} -- is, for $G>4$, consistent with its UCL analog.} (US) members by \citet{2022NatAs...6...89M} pushing completeness down to $\sim 10 \mjup$. 2MASS $J$ magnitudes were converted into Gaia $G$ magnitudes based again on BHAC15 isochrones at 15 Myr, yielding the $\zeta_{\text{US}}(G)$ PDF of the sample; a new normalized $\zeta(G)$ could then be built by combining the Gaia-based UCL list and the US sample, setting a sharp transition between the former and the latter at the dimmest magnitude where they intersect ($\hat{G}=16.8$ mag): above this value, the two distributions start to significantly differ due to Gaia incompleteness. The number of unseen sources $n_U$ recovered in this way amounts to $\approx 700$ ($\sim 18\%$), and their PDF is given by:

\begin{equation}
n_U(G) =
\begin{cases}
0 & G<\hat{G}\\
N_{\text{UCL}} \cdot [\zeta_{\text{US}}(G))-\zeta_{\text{Gaia}}(G)] & G \geq \hat{G}
\end{cases}\,.    
\end{equation}

In order to compute the probability that a phantom or an unseen star is hidden by a BEAST star, we consider as a typical BEAST star an object as bright in the apparent G-band as the mean of the sample ($G=5.29$); for a contrast $\Delta G \approx 8$ mag, the effective shading separation $s_{\text{eff},B}$ of this star starts being larger than the half-edge (5.5") of IRDIS FOV ($A_{\text{IRD}} = 11" \times 11"$); we therefore impose $s_{\text{eff},B}(\Delta G) = \text{inf}(s_{\text{eff},B}(\Delta G),\sqrt{A_{\text{IRD}}/\pi}$). The differential probability associated to the event as a function of $G$ is given by:
\begin{equation}
f(G) = f_P(G) + f_U(G) = \frac{n_P(G) \cdot \pi s_{\text{eff},B}^2(G)}{A_s(G)}+ \frac{n_U(G) \cdot A_{\text{IRD}}} {A_{\text{UCL}}}
\end{equation}
where the second term takes into account that unseen sources should be spread over the entire UCL. Integration of $f(G)$ yields $p=\int_{G=5.29}^{G=25} f(G) dG = 2.8 \cdot 10^{-5}$. The false alarm probability associated to the event of finding at least one such object across the whole survey, having completed until now the observations of 47 stars, is equal to:
\begin{equation}
    \text{FAP} = 1-\binom{47}{0}(1-p)^{47} = 1.3 \times 10^{-3}.
\end{equation}

In order to evaluate the impact of the assumption of a constant age for US -- which is instead known to have experienced a long-lasting star formation history ranging between 15 and 5 Myr ago \citep{squicciarini21,ratzenbock23} --, the conversion of $J$ magnitudes into $G$ magnitudes was repeated by supposing a constant age $5$ Myr. The resulting $\text{FAP}\approx 1.4 \cdot 10^{-3}$ provides robust evidence for the independence of the result on model assumptions, firmly allowing us to exclude the alternative scenario in favor of the one positing a single quadruple system.

\section{Characterization of Cb}
\label{Appendix_Cb_mass}

The derivation of a photometric mass estimate for HIP 81208 Cb is mediated by \textsc{madys} \citep{madys}, as in previous BEAST publications \citep[]{mu2_sco, Viswanath_hip81208_tmp}. After averaging $K1$ and $K2$ contrasts derived by \verb+PACO+ over the two epochs, the conversion of those contrasts into calibrated apparent magnitudes was operated by means of the 2MASS $K_s$ magnitude of the primary. Being HIP 81208 A classified as a B9V star, the impact of the approximation $K_{s,A} \approx K1_{A} \approx K2_{A}$ is well within the photometric error budget \citep{mamajek13}.

Interstellar reddening towards HIP 81208 is known to be rather small \citep{Viswanath_hip81208_tmp} and translating, in the $K_s$ band, to a negligible $A_{K_s} = 0.003 \pm 0.006$ adopting $A_{K_s}/\ebv = 0.306$ \citep{yuan13}. Likewise, the adopted parallax and age estimates reflect those used in the \citet{Viswanath_hip81208_tmp} paper.

\noindent Having obtained absolute magnitudes for the object:
\begin{align*}
K1 = 9.90 \pm 0.04~\text{mag}, \\
K2 = 9.48 \pm 0.04~\text{mag},
\end{align*}

we built a set $\mathcal{M}$ of substellar evolutionary models that are adequate for the age and mass range of interest, while providing at once synthetic SPHERE magnitudes: such a set, to which observed magnitudes were compared, comprehends the Ames-Dusty models \citep{ames_models} and the BT-Settl models \citep{bt_settl_models}. Details on the fitting algorithm, encompassing all sources of uncertainty within a Monte Carlo framework, can be found in \citep{madys}. The output of each model $i \in \mathcal{M}$ corresponds to a triplet ($M_{min,i},M_{opt,i},M_{max,i}$) equal to the ($16^{th},50^{th},84^{th}$) percentiles of the posterior mass distribution; the two outputs were averaged in the following manner:
\begin{align}
M_{min} &= \inf_{i \in \mathcal{M}} (\{M_{min,i}\}) \label{eq:final_mass1}\\
M_{opt} &= \underset{i \in \mathcal{M}}{\text{mean}} (\{M_{opt,i}\}) \label{eq:final_mass2}\\
M_{max} &= \sup_{i \in \mathcal{M}} (\{M_{+,i}\}) \label{eq:final_mass3}
\end{align}
with the goal of embedding theoretical uncertainties onto the final estimate. 

The posterior mass distribution returned by each model can be easily translated into the posterior distribution of any astrophysical parameter of interest provided by the isochrone grids. We were therefore able to derive in a self-consistent way (using similar equations to Eq.~\ref{eq:final_mass1}-\ref{eq:final_mass3})) the best-fit estimates for effective temperature, surface gravity and bolometric luminosity. Likewise, we also computed synthetic 2MASS $H$ and $K$ magnitudes as a helpful first-guess estimate for follow-up studies:
\begin{align*}
H &= 10.28\pm0.07~\text{mag}, \\
K_s &= 9.74 \pm 0.04~\text{mag}.
\end{align*}

We highlight that independent mass determinations were derived {\textit{a posteriori}}, and used as a control sample, starting from the best-fit $\log{L/\Lsun}$ through the recent ATMO2020 \citep{atmo2020} and Sonora Bobcat \citep{sonora} grids\footnote{These grids were not included in $\mathcal{M}$ because they are currently not equipped with SPHERE filters.}; these best-fit masses of $14.51^{+0.16}_{0.15} \mjup$ and $14.44^{+0.15}_{0.14} \mjup$, respectively, are consistent with our best-fit mass estimates. Nonetheless, as already mentioned in Sect.~\ref{sec:Cb_astro_parameters}, we are not able to exclude the possibility of unaccounted systematic effects that are common to all the adopted models.

\section{Orbital fit of Cb: corner plot}
\label{Appendix_corner_plot}

Given the small separation between C and Cb, most sources of systematic error are either canceled out (centering error) or significantly decreased (platescale and True North error), enabling an accurate determination of their relative separation at both epochs (Appendix~\ref{Appendix_PSF}).

Starting from the relative C-Cb astrometry as measured by PSF subtraction and their best-fit masses (cp. Table~\ref{tab:system_architecture}), we ran an MCMC code based on the \verb+emcee+ code \citep{2013PASP..125..306F} in order to derive the orbital parameters of Cb's orbit around C. 

The input parameters for the MCMC are a logarithmically uniform prior for semimajor axis ($a \in [0, 80]$ au) and an eccentricity $e \in [0, 0.4]$. 

\begin{figure*}[t!]
    \centering
    \includegraphics[width=\textwidth]{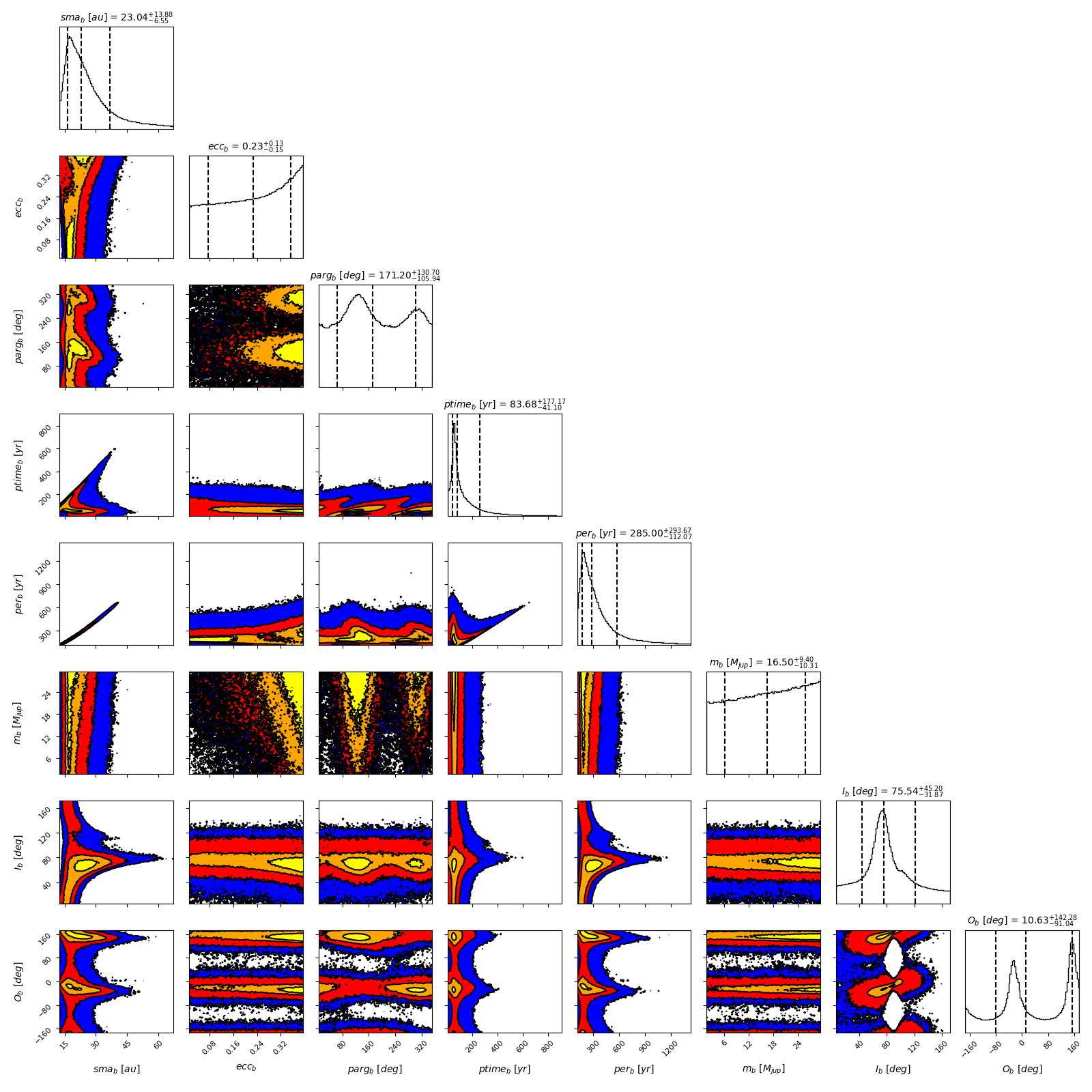}
    \caption{Corner plot of the orbital solution for Cb. Reading from left to right, the acronyms stand for (from top to bottom) {\textit{semimajor axis}}, {\textit{eccentricity}}, {\textit{argument of periapsis}}, \textit{periapsis time}, {\textit{orbital period}}, {\textit{Cb mass}}, {\textit{inclination}}, and {\textit{longitude of the ascending node}}.}
    \label{fig:corner_plot_Cb}
\end{figure*}

The posterior distribution for the orbital parameters of Cb derived in this work is provided in Fig.~\ref{fig:corner_plot_Cb}.

\end{appendix}
\end{document}